\documentclass{aa}  

\usepackage{graphicx}
\usepackage{txfonts}
\usepackage{footnote}
\makesavenoteenv{tabular}
\usepackage{graphicx}
\usepackage{float}
\usepackage{stfloats}
\usepackage{lipsum,caption,graphicx}
\usepackage[dvipsnames]{xcolor}
\usepackage{color, colortbl}
\usepackage{multicol}
\usepackage[toc]{appendix}
\usepackage[normalem]{ulem}
\definecolor{darkred}{RGB}{148,17,0}
\definecolor{lightred}{RGB}{231,211,207}
\definecolor{darkblue}{RGB}{0,84,147}
\definecolor{lightblue}{RGB}{210,221,232}
\definecolor{darkgreen}{RGB}{30,123,30}
\definecolor{lightgreen}{RGB}{214,228,228}
\definecolor{orange}{RGB}{255,140,0}
\usepackage{gensymb}
\usepackage{threeparttable}
\usepackage{textcomp}

\usepackage{tikz}
\usetikzlibrary{fit,matrix, shapes, arrows, positioning,calc,scopes,fadings,angles,arrows.meta,arrows.meta,angles,quotes,calc}
\usetikzlibrary{decorations.pathmorphing}
\usetikzlibrary{decorations.pathreplacing,calligraphy}
\usetikzlibrary{decorations.markings}
\usetikzlibrary{decorations.text}

\begin{document}

   \title{Evolution of pits at the surface of 67P/Churyumov-Gerasimenko}

\author{ 
Selma Benseguane\inst{1}
\and
Aur\'elie Guilbert-Lepoutre\inst{1}
\and
J\'er\'emie Lasue\inst{2}
\and\\
S\'ebastien Besse\inst{3}
\and
C\'edric Leyrat\inst{4}
\and
Arnaud Beth\inst{5}
\and\\
Marc Costa Sitj\`a\inst{6}
\and
Bj\"orn Grieger\inst{3}
\and
Maria Teresa Capria\inst{7}
}

\institute{LGL-TPE, CNRS, Université Lyon, UCBL, ENSL, F-69622 Villeurbanne, France\\
          \email{selma.benseguane@univ-lyon1.fr}
\and
IRAP, Université de Toulouse, CNRS, CNES, UPS, F-31400 Toulouse, France
\and
Aurora Technology B.V. for the European Space Agency, ESAC, 28692 Villanueva de la Canada, Madrid, Spain
\and
LESIA, Observatoire de Paris, CNRS, Sorbonne Univ., Univ. Paris-Diderot, Meudon, France
\and
Department of Physics, Ume\aa\ University, 901 87 Ume\aa\ , Sweden
\and
Rhea System for the European Space Agency, ESAC, 28692 Villanueva de la Canada, Madrid, Spain
\and
Istituto di Astrofisica e Planetologia Spaziali (IAPS), INAF, I-00133 Roma, Italy
}

\titlerunning{Evolution of pits at the surface of 67P}
\authorrunning{Benseguane et al.}

\date{Received ; accepted }


\abstract
{The observation of pits at the surface of comets offers the opportunity to take a glimpse into the properties and the mechanisms that shape a nucleus through cometary activity. If the origin of these pits is still a matter of debate, multiple studies have recently suggested that known phase transitions (such as volatile sublimation or amorphous water ice crystallization) alone could not have carved these morphological features on the surface of 67P/Churyumov-Gerasimenko (hereafter 67P).}
{We want to understand how the progressive modification of 67P's surface due to cometary activity might have affected the characteristics of pits and alcoves. In particular, we aim to understand whether signatures of the formation mechanism of these surface morphological features can still be identified.}
{To quantify the amount of erosion sustained at the surface of 67P since it arrived on its currently observed orbit, we selected 380 facets of a medium-resolution shape model of the nucleus, sampling 30 pits and alcoves across the surface. We computed the surface energy balance with a high temporal resolution, including shadowing and self-heating contributions. We then applied a thermal evolution model to assess the amount of erosion sustained after ten orbital revolutions under current illumination conditions.}
{We find that the maximum erosion sustained after ten orbital revolutions is on the order of 80~m, for facets located in the southern hemisphere. We thus confirm that progressive erosion cannot form pits and alcoves, as local erosion is much lower than their observed depth and diameter. We find that plateaus tend to erode more than bottoms, especially for the deepest depressions, and that some differential erosion can affect their morphology. As a general rule, our results suggest that sharp morphological features tend to be erased by progressive erosion.}
{This study supports the assumption that deep circular pits, such as Seth\_01, are the least processed morphological features at the surface of 67P, or the best preserved since their formation.}


   \maketitle


\section{Introduction}

Comets are among the least processed remnants of the early stages of our planetary system. The study of comets thus provides critical information to help us better understand the physical processes that lead to planet formation, and the early material that formed the protoplanets \citep{festou2004,cochran2015}.
Jupiter-family comets (JFCs) are a subpopulation of comets, with short-period orbits dominated by the gravitational influence of Jupiter. They are thought to originate from the Kuiper Belt and the scattered disk \citep{brasser2013}, where they got destabilized owing to interactions with Neptune. They evolved through the giant-planet region and toward the inner solar system where they are observed nowadays \citep{levison1997,disisto2009,nesvorny2017}.
Cometary activity starts beyond the orbits of Jupiter and Saturn for long-period comets \citep[][]{meech2017,jewitt2017,hui2017,hui2019,yang2021,farnham2021} and for Centaurs, which are the precursors of JFCs \citep{jewitt2009,lin2014,epifani2017,epifani2018,steckloff2020,delafuentemarcos2021}.  
This implies that JFCs, which have been studied so far by space missions, mostly have evolved surfaces \citep[e.g.,][]{gkotsinas2022}. 
In this framework, the European Space Agency (ESA)'s \emph{Rosetta} mission aimed to study how a comet's surface might be modified through cometary activity. Indeed, by understanding the physical processes that currently reshape the nucleus, we might reconstruct the properties that it would have had at the time of its formation.\\

Significant geological heterogeneity was observed at the surface of 67P/Churyumov-Gerasimenko (hereafter 67P). In addition to the presence of terraces, strata, fractures \citep{massironi2015}, goose-bump features \citep{sierks2015}, and wind-tail-like features \citep{el-maarry2019}, the observation of surface depressions, linked with cometary activity, thus offered the opportunity to look into the characteristics of the subsurface and the thermophysical processes actively modifying them \citep{sierks2015,vincent2015a}.
Two main types of depressions can be distinguished on the surface of 67P based on their dimensions, that is to say their diameter and their depth (mainly the latter): shallow depressions and deep depressions, as we detail below.\\

First, shallow depressions of only a few meters in depth are generally observed on smooth terrains, in many regions across the nucleus. These might be seasonal depressions shaped during perihelion passages, reported to be mainly driven by sublimation activity in the current orbits of the comet. For instance, \citet{vincent2016} and \citet{el-maarry2017} propose that the surface was reshaped via scarp retreat, and \citet{pajola2016} infer that these shallow structures could indicate a future cliff collapse. \citet{groussin2015} and \citet{bouquety2021a} suggest that they could be seasonal structures shaped by progressive erosion, induced by activity sustained close to the perihelion approach.
\citet{bouquety2021} named these depressions cometary thermokarst depressions due to their morphometrical analogy with thermokarstic lakes on Earth and scalloped terrain on Mars. These depressions will not be further studied in this work.\\

\begin{figure}[h!]
    \centering
    \includegraphics[width=\columnwidth]{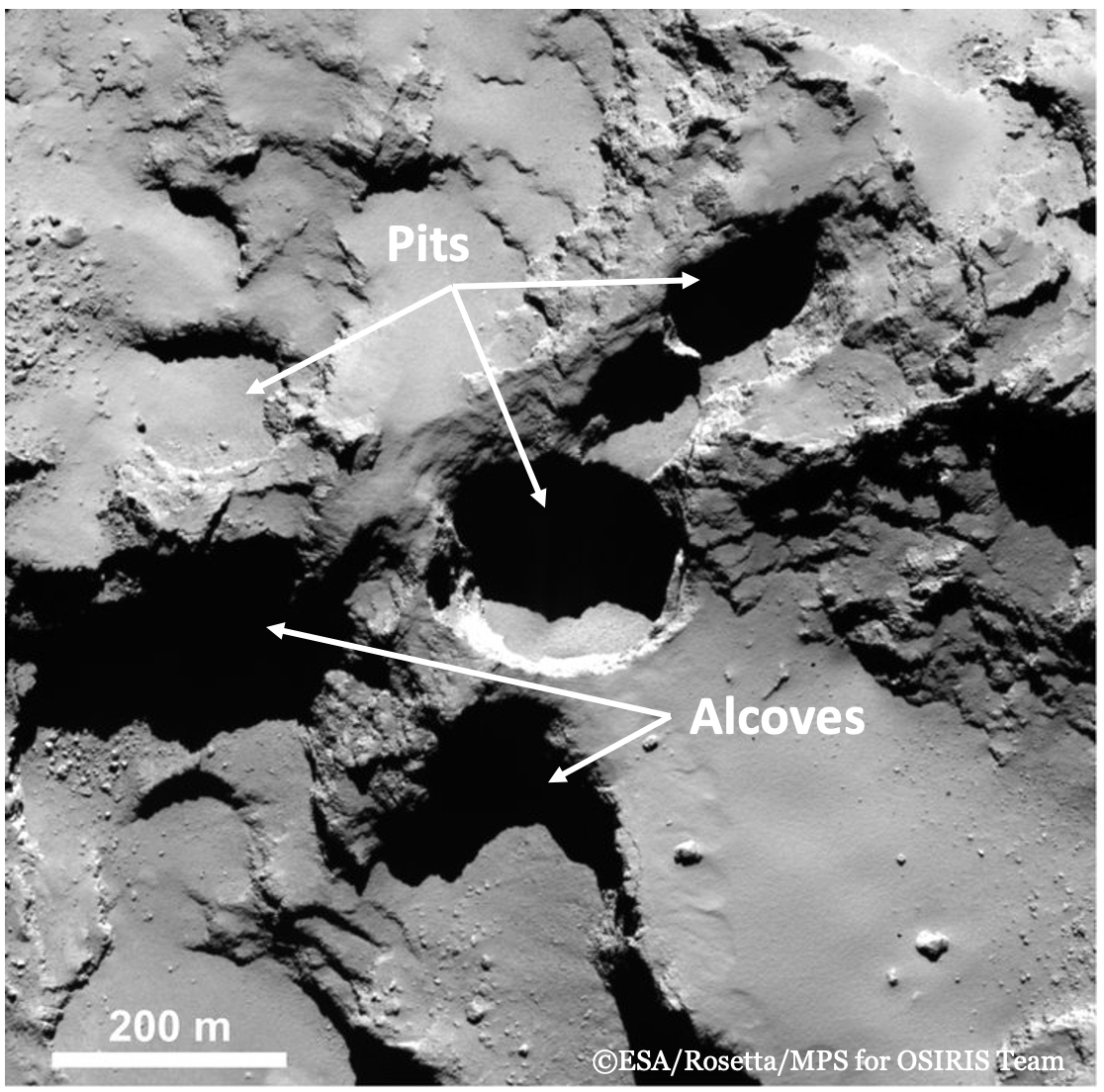} 
    \caption[width=\columnwidth]{Image from OSIRIS/NAC of a part of the Seth region on which we illustrate the type of depressions we study: pits and alcoves \citep[half circular-pits,][]{vincent2015a}.}  
    \label{pits_nac}
\end{figure}

In this study, we are interested in the second kind of surface depressions, characterized by steep walls with depths of tens to hundreds of meters \citep{massironi2014,el-maarry2015,thomas2015a,el-maarry2019}. 
These larger-scale structures include pits, as well as cliffs or alcoves (see Fig.~\ref{pits_nac} retrieved from \emph{Rosetta}/OSIRIS' NAC (Narrow Angle Camera)).
They are mostly present on 67P's northern hemisphere and are generally concentrated in some regions. For instance, the Maftet geological unit displays irregular-shaped pits of 10 to 20~m deep and 100 to 150~m in diameter \citep{thomas2015}.
The Seth region is dominated by multiple series of circular, flat-floored pits \citep{besse2015} and contains a pit chain similar to the one observed on Ma'at \citep[][see Fig.~\ref{pits_nac}]{thomas2015}. It also contains cliffs that are tens to hundreds of meters high, which are also observed in Hathor \citep{el-maarry2015}. 
Furthermore, \citet{vincent2015a} report the detection of cometary activity in the form of localized dust jets in some of these pits. Additionally, they note that active pits have a high depth-to-diameter ratio compared to inactive ones \citep{besse2015}. 
Our study thus specifically considers the following pits: Seth\_01, Seth\_02, Seth\_03, Seth\_04, Seth\_05, Seth\_06, Ma'at\_01, Ma'at\_02, Ash\_03, Ash\_04, Ash\_05, and Ash\_06 \citep[see Fig.~1 from][]{vincent2015a}. We also include additional pits, with similar geomorphological characteristics to the ones studied in \citep{vincent2015a}. 
We further study cliffs or alcoves, as these might be construed as deteriorated pits \citep{vincent2015a}.

As a result, our study focuses on features of at least a few tens of meters in depth, and a few hundreds of meters in diameter: the smallest depth and diameter are 35~m and 130~m, respectively.
In the rest of the paper, we indifferently use the term ``pit'' for the sake of simplicity.\\


Such pits have been observed on most comets directly studied by space missions, for example 19P/Borelly \citep[seen by \emph{Deep Space 1},][]{soderblom2002}, 81P/Wild~2 \citep[seen by \emph{Stardust},][]{brownlee2004}, 9P/Tempel~1 \citep[seen by \emph{Deep Impact} and \emph{ Stardust-NExT},][]{belton2013}, and 103P/Hartley~2 \citep[seen by \emph{EPOXI},][]{syal2013}.
The mechanism at the origin of these structures is still a matter of debate \citep{brownlee2004, belton2009, belton2013, thomas2013}. 
\citet{holsapple2007} and \citet{vincent2015a} argue that impacts on cometary surfaces are expected to produce features with a morphology distinct from these observed pits, and thus they should be a signature of some process related to cometary activity rather than the result of collisions.
\citet{vincent2015a}, \citet{kossacki2018} and \citet{leliwa-kopystynski2018} propose the formation of pits by sinkhole collapse due to subsurface cavities, either primitive or formed as a result of subsurface depletion of volatiles by ice sublimation. \citet{massironi2014} argue that a sublimation process can lead to slope retreats and material ablation at the pit's location, while \citet{thomas2015a} argue that mechanisms such as ice sublimation or sinkhole collapse would not likely lead to the material structure giving rise to the quasi-circular aspect of pits.

\citet{mousis2015} explored the possibility such structures forming due to phase transitions (i.e., sublimation, amorphous water ice crystallization, and clathrate destabilization). They showed that the time required to produce features of the spatial scale observed by \emph{Rosetta} on the surface of 67P is long, on the order of a thousand years or more. 
\citet{guilbert-lepoutre2016} further showed that it is very unlikely that pits form with the current illumination conditions, as no known mechanism could carve the surface to form pits with a depth of $\sim$200 m, and a diameter ranging from 100 to 300~m over short timescales.
Moreover, since 67P's previous orbits had perihelion distances farther away from the Sun \citep{maquet2015}, it is unlikely that quasi-circular pits were formed by the progressive effect of one phase transition.
Because the distribution law of the pits' size frequency is similar at the surface of 67P, 9P/Tempel~1, and 81P/Wild~2, \citet{ip2016} suggested that they might have been formed with the same mechanism operating on many JFCs. Comparing the orbital history of those comets, they inferred that such processes might have carved pits before these comets entered the inner solar system.

With these arguments in mind, we want to understand how the progressive modification of 67P's surface due to cometary activity might have affected the characteristics of these depressions. In particular, we are interested in understanding whether signatures of the formation mechanism at the origin of pits can still be found. Our goal is thus to quantify the amount of erosion sustained by pits at the surface of 67P, under the current illumination conditions that periodic, daily, and seasonal cycles entail. In Sect. \ref{sec:methods} we present the surface energy model and the thermophysical evolution model used to address this issue. We present the results of the influence of several key parameters in our study in Sect. \ref{initial-params}, and the results of the thermal simulations in Sect. \ref{sec:thermal-simu}. Finally, we discuss our results in Sect. \ref{sec:discussion}.

\section{Methods} \label{sec:methods}

Our study follows the work of \citet{mousis2015} and \citet{guilbert-lepoutre2016}, who constrained the thermal evolution of 67P's subsurface over the recent past. They studied the possible formation of pits in general, averaging the energy input across 67P's surface. To go beyond these studies, our goal is to quantify how the energy received locally, at a small scale on the surface, translates into phase transitions. We aim to quantify the extent of these phase transitions over multiple perihelion passages. Eventually, we want to assess whether these can be the origin of the formation of pits, or their evolution into structures with the spatial scale observed by \emph{Rosetta}.
To do so, we used a Stereo-PhotoGrammetric (SPG) shape model of 67P's nucleus \citep[][see Sect. \ref{sec:shapemodel} for details]{preusker2017}. The spatial resolution was chosen to provide several facets for each geometric portion of a pit (i.e., the bottom, the cliffs, and the plateau surrounding it). The energy received by each facet was computed and used as the surface boundary condition of a thermal evolution model. Each step of this method is detailed in the following sections.

\subsection{Thermophysical evolution model}

\subsubsection{Main equations}

A thermophysical evolution model was applied to each facet of this SPG model. 
The following aspects needed to be taken into consideration when choosing our numerical scheme: 

Each facet gets its own boundary condition at the surface. Therefore, a 1D thermal evolution model is the best option, further justified by the results of \citet{macher2019}, who found that temperature differences at the surface of 67P between a 1D and a 3D thermal simulation (i.e., accounting for lateral heat fluxes) amount to only $\sim$ 0.1\%.

Physical processes included in the model should be standard to thermal evolution models developed over the past few decades \citep{prialnik2004,huebner2006}: heat and gas diffusion, phase transitions for volatile species, drag of dust particles by the escaping vapor phase, and the formation of a dust mantle at the surface.

Finally, thermal evolution models necessarily rely on a number of free or poorly constrained thermophysical characteristics. Exploring the free-parameter-space ought to be rapid from a computational point of view, so as to provide an insight into the robustness of our results. Therefore, simple expressions of thermophysical characteristics such as the thermal conductivity for example are preferred to complex ones, as these would introduce additional parameters.\\ 

With these considerations in mind, we chose the 1D scheme as the basis of multiple models. We refer the reader to the works by \citep{desanctis2005,desanctis2010,lasue2008} for details on the model, but we provide the main equations below for clarity. More sophisticated models exist to study the thermal evolution of cometary nuclei, considering several dimensions \citep{guilbert-lepoutre2016}, and refined descriptions of each thermophysical parameter \citep{davidsson2021a}. However, because our purpose is to understand in detail the influence of the energy input, modulated by local topography and the global morphology of 67P's nucleus, we need to keep our thermal evolution model relatively simple. Our model thus solves the heat diffusion equation:
\begin{equation}
\rho_{bulk} c \frac{\partial T}{\partial t} = \mathrm{div} \left(\kappa~\overrightarrow{\mathrm{grad}}~T \right) + \mathcal{S}
\label{e_conserv},
\end{equation}
where $\rho_{bulk}$ [kg~m$^{-3}$] is the material's bulk density, $c$ [J~kg$^{-1}$~K$^{-1}$] its heat capacity, $\kappa$ [W~m$^{-1}$~K$^{-1}$] its thermal conductivity, and $\mathcal{S}=\mathcal{Q}_{cr}+\sum_{\alpha}\mathcal{Q}_{\alpha}$ the energy sources and sinks. 
For our study, we take into account two such heat sources and sinks. The first one is the energy released upon crystallization of amorphous water ice, assuming it is exothermic:
\begin{equation}
\mathcal{Q}_{cr} = \lambda (T)~ \varrho _{am}~ \Delta H_{ac},
\end{equation}
where $\varrho_{am}$ [kg~m$^{-3}$] is the mass of amorphous water ice per unit volume. The phase transition releases a latent heat $\Delta H_{ac}$ = 9$\times$10$^{4}$~J~kg$^{-1}$ \citep{klinger1981} at a rate of $\lambda (T) = 1.05 \times 10^{13} ~ e^{-5370/T}$~s$^{-1}$, determined by \citet{schmitt1989}.
The second is the energy loss (or gain) due to sublimation (or recondensation) of different ices present in the solid material. We assume a simple composition of dust and ice, with H$_2$O, CO, and CO$_2$ present as pure compounds in the initial icy matrix. For each ice, we have:
\begin{equation}
\mathcal{Q}_{\alpha}=-\psi~\Delta H_{\alpha}~q_{\alpha},
\end{equation}
where $\psi$ is the porosity; $\Delta H_{\alpha}$ [J~kg$^{-1}$] is the latent heat of sublimation of species $\alpha$ (H$_2$O, CO, or CO$_2$); and $q_{\alpha}$  is the related gas source term, which is obtained through mass conservation equations.

Assuming that sublimation of amorphous water ice is negligible, because the phase transition to crystalline water ice occurs first at lower temperatures, the set of mass balance equations may be written as:
\begin{equation}
\frac{\partial \varrho_{am}}{\partial t} = -\lambda(T)~\varrho_{am},
\end{equation}
\begin{equation}
\frac{\partial \varrho_{cr}}{\partial t} = \lambda(T)~\varrho_{am} - q_{H_2O},
\end{equation}
\begin{equation}
\frac{\partial \tilde{\varrho}_{\alpha}}{\partial t}  + \mathrm{div}~\overrightarrow{\phi_{\alpha}}= q_{\alpha},
\end{equation}
where $\varrho_{am}$ and $\varrho_{cr}$ [kg~m$^{-3}$] are the mass per unit volume of amorphous and crystalline water ice, respectively, and $\tilde{\varrho}_{\alpha}$ [kg~m$^{-3}$] is the mass per unit volume of each gas species. We assume that the vapor and the solid phases are in local thermodynamic equilibrium, and the vapor phase behaves as an ideal gas (i.e., no interaction between species). Each gas flux $\overrightarrow{\phi_{\alpha}}$ can thus be written separately, as:
\begin{equation}
\overrightarrow{\phi_{\alpha}} = -\mathcal{G}_{\alpha}~ \overrightarrow{\mathrm{grad}}~\mathcal{P}_{\alpha},    
\end{equation}
with $\mathcal{P}_{\alpha}$ being the partial pressure of each species, and $\mathcal{G}_{\alpha}$ being a gas diffusion coefficient that generally depends on the structural parameters of the solid matrix (such as the porosity, the size of pores, or the tortuosity), and the temperature (see \citealt{prialnik2004} or \citealt{huebner2006} for details). 
For each volatile species, the gas source term can thus be written as:
\begin{equation}
q_{\alpha} = \frac{1}{\mathcal{R}T} \frac{\partial\mathcal{P}_{\alpha}}{\partial t} - \mathrm{div}\left( \mathcal{G}_{\alpha}~ \overrightarrow{\mathrm{grad}}~\mathcal{P}_{\alpha} \right) 
\label{m_conserv_new},
\end{equation}
where $\mathcal{R}$ is the ideal gas constant.

\subsubsection{Initial parameters}

The composition and internal structure of cometary nuclei are generally poorly known. The \emph{Rosetta} mission has, however, provided some crucial measurements for 67P, which are used as constraints in our model whenever possible. All the parameters included in the thermal evolution model have, nonetheless, not been measured, and we therefore make standard assumptions regarding the values of the unknown parameters (see Table \ref{init_params}, and \cite{huebner2006} for details).

\begin{table*}[ht]
    \centering
    \caption{Initial parameters for the thermal evolution model.}   
    \begin{tabular}
    [width=\textwidth]
    {lllrl}
    \hline
    Parameter && Symbol & Value & Unit \\
    \hline \hline
    Bond albedo &&$\mathcal{A}$, $\mathcal{A}_R$, $\mathcal{A}_T$ &0.06 &-  \\ 
    Emissivity &&$\varepsilon$ &0.95 &- \\
    Initial temperature &&T$_i$ &30 &K \\
    Hertz factor &&$f_H$ &0.005 &- \\
    Pore radius &&$r_{pore}$ &10$^{-4}$ &m \\    
    Porosity &&$\psi$ &75 &\% \\
    
    Mass per unit volume &Dust  &$\rho_d$ &1000  &kg m$^{-3}$ \\
    &H$_2$O ice  &$\rho_{am}$, $\rho_{cr}$ &917 &kg m$^{-3}$  \\
    &CO ice  &$\rho_{H_2O}$ &1250  &kg m$^{-3}$ \\
    &CO$_2$ ice  &$\rho_{H_2O}$ &1977 &kg m$^{-3}$  \\
    
    Mass fraction & Dust/H$_2$O  &$X_d/X_{H_2O}$ &1 &- \\
    &CO/H$_2$O &$X_{CO}/X_{H_2O}$ &0 &- \\
    &CO$_2$/H$_2$O &$X_{CO_2}/X_{H_2O}$ &0 &- \\
    
    Thermal conductivity & Dust  &$\kappa_d$ &3 &W m$^{-1}$ K$^{-1}$ \\
    &Crystalline water ice  &$\kappa_{cr}$ &567/T &W m$^{-1}$ K$^{-1}$ \\
    &Amorphous water ice  &$\kappa_{am}$ &2.34 $\times$ 10$^{-3}$ T + 2.8 $\times$ 10$^{-2}$ &W m$^{-1}$ K$^{-1}$ \\

    Heat capacity & Dust &$c_d$ &1300  & J kg$^{-1}$ K$^{-1}$ \\  
    &H$_2$O ice &$c_{H_2O}$ &1610 & J kg$^{-1}$ K$^{-1}$ \\
    &CO ice &$c_{CO}$ &2010 & J kg$^{-1}$ K$^{-1}$ \\
    &CO$_2$ ice&$c_{CO_2}$ &1610 &J kg$^{-1}$ K$^{-1}$ \\
    
    Latent heat of sublimation &H$_2$O &$\Delta H_{H_2O}$ &2.83 $\times$ 10$^6$  &J kg$^{-1}$ \\ 
    &CO &$\Delta H_{CO}$ &0.29 $\times$ 10$^6$ &J kg$^{-1}$ \\  
    &CO$_2$ &$\Delta H_{CO_2}$ &0.58 $\times$ 10$^6$ &J kg$^{-1}$ \\ 
    \hline
    \end{tabular}
    \label{init_params}
\end{table*}

We note that some thermophysical parameters depend on one another. For example, the bulk density of a cometary nucleus can be written as:
\begin{equation}\label{bulkdensity}
\rho _{bulk} = (1 - \psi) \left( \sum_{i} \frac{X_i}{\rho_i}  \right) ^{-1},
\end{equation}
where $\psi$ is the porosity, $X_i$ is the mass fraction of each individual component in the cometary material mixture, and $\rho_i$ [kg m$^{-3}$] is the corresponding solid density of each constituent. 
If we use the bulk density of 533$\pm$6~kg~m$^{-3}$ measured by \emph{Rosetta} for 67P's nucleus \citep[][]{patzold2016}, 
a degeneracy remains between the composition and the porosity to obtain this value. Moreover, we study processes affecting the $\sim$100\,m-surface layer, which might not have  the same properties as the bulk of the nucleus deep inside. In our model, we thus chose the porosity and dust-to-ice mass ratio, which gives an associated bulk density on the same order of the observed bulk density.
The composition and porosity also influence the value of thermal characteristics such as the thermal conductivity (i.e., $\kappa = f_{\psi}~f_H~\frac{ \sum_i M_i \kappa_i}{ \sum_i M_i }$, where $f_{\psi}$ and $f_H$ are respectively the correction factors to account for the porosity and the reduced contact between solid grains, also known as the Hertz factor, $M_i$ is the mass per unit volume of each constituent $i$, and $\kappa_i$ is their respective thermal conductivity) or the heat capacity (i.e., $c = \frac{ \sum_i M_i c_i}{ \sum_i M_i}$, where $M_i$ is the mass per unit volume of each constituent $i$, and $c_i$ is their respective heat capacity). Thus, we test several values of the most crucial characteristics, in order to assess their influence on the outcome of thermal evolution simulations.  
These parameters are: the initial porosity of the cometary material (Sect. \ref{sec:porosity}); the dust-to-ice mass ratio (Sect. \ref{sec:dusttoice}); the abundances of CO and CO$_2$ (Sect. \ref{sec:abundance}); and the thickness of the dust mantle at the surface (Sect. \ref{sec:mantle}).

\subsection{Boundary conditions}

\subsubsection{Shape model} \label{sec:shapemodel}

A shape model of 67P's nucleus was reconstructed using the SPG technique on high-resolution images taken by the \emph{Rosetta}/OSIRIS instrument \citep{preusker2015}. The latest SHAP7 SPG shape model reaches a very high spatial resolution, with 44 million facets, reconstructed from 1500 OSIRIS' NAC images \citep{preusker2017}\footnote{http://europlanet.dlr.de/Rosetta/}. From this model, several lower-resolution models were derived: in this study, we use an SPG shape model composed of 124,938 facets
\footnote{http://comsim.esac.esa.int/rossim/SHAPE\_MODEL\_DRAFTS/\\SHAP7\_8/SPG/shap7\_model\_info.asc.}. With this model, the typical average distance between two nodes of a facet is $\sim$20~m. Local topography and roughness at smaller scale thus cannot be accounted for in our work.

\subsubsection{Energy balance at the surface}

The energy and mass conservation equations need to be constrained by boundary conditions. The energy equilibrium boundary condition at the surface is given for each facet by:
\begin{equation}
        (1-\mathcal{A}_R)~ \mathcal{E}= \varepsilon \sigma T^4 + \kappa \frac{\partial T}{\partial r} +  \sum_{\alpha}f_{\alpha} \Delta H_{\alpha} Q_{\alpha}
        \label{surf_bound},
\end{equation}
where $\mathcal{A}_R$ is the Bond albedo of the facet for which we compute the energy balance, $\varepsilon$ is the emissivity, $\sigma$ is the Stefan–Boltzmann constant, and $T$ [K] is the surface equilibrium temperature. We allow for the presence of volatile species at the surface, so that sublimation is possible: $f_{\alpha}$ represents the fraction of the facet's surface covered by these ices, and $Q_{\alpha}$ [kg~m$^{-2}$~s$^{-1}$] is the corresponding sublimation rate. 
Finally, $\mathcal{E} = E_{\odot} + E_{IR} + E_{VIS}$ is the total energy flux received by a given facet, which takes into account the contributions (detailed below) from direct insolation $E_{\odot}$, and hence shadowing effects due to the complex global morphology of 67P's nucleus, and from self-heating $E_{IR} + E_{VIS}$, namely the energy flux received by reflection and emission from neighboring facets in the visible and infrared, respectively.

Direct insolation is given by:
\begin{equation}
    E_{\odot} = \frac{F_{\odot}}{r_H^2}~ \cos{\xi},
\end{equation}
where $F_{\odot}$ [W~m$^{-2}$] is the solar flux at 1~au. The heliocentric distance $r_H$ [au] and the local zenith angle $\xi$ both vary with time. For each time step, we first retrieve the coordinates of the subsolar point using SPICE database kernels, which contain the information on both the rotation state of 67P's nucleus and its orbital parameters. Then, the insolation geometry for each facet is computed with respect to the subsolar point's coordinates. For facets located on the night side of the nucleus, we apply the following criterion: if $\cos{\xi}$~<~0 then $E_{\odot}$~=~0. To assess which facets are located in the shadow of global or local topographic features, we project the nodes of the shape model on a 2D plane normal to the zenith direction of the subsolar point. For each node, we compute its projected position along the normal direction, and test whether it is below an other facet: this node is then considered shadowed. When one of the three nodes of a facet is shadowed, we consider that the whole facet is shadowed. 

Given the complex morphology of 67P's nucleus, observed both on a global scale (two lobes) and on a local scale \citep[e.g.,][]{el-maarry2015}, self-heating -- the energy flux received by reflection or emission from neighboring facets -- might be a significant additional source of energy. It is composed of two contributions, one from visible radiations reflected by mutually facing facets, $E_{VIS}$, and one from their thermal infrared emissions, $E_{IR}$.
We note that the contribution from infrared radiations reflected by mutually facing facets is not taken into account, because it is always negligible compared to the other two self-heating contributions. 
The relative influence of self-heating thus depends on how facets from the shape model see each other: the self-viewing geometry is a function of the orientation of the mutual facing facets, both emitting and receiving. For the facet of interest, for which the energy balance is being computed, the visible contribution of self-heating can be written as:
\begin{equation}
E_{VIS} = \sum_T ~~ \mathcal{A}_T~ \frac{F_{\odot}}{r_H^2} \cos{\xi_{T}} ~\frac{S_T}{\pi} ~\frac{\cos{\zeta_{T}}~\cos{\zeta_{R}}}{\delta^2_T} 
\label{e_vis},
\end{equation}
where $\mathcal{A}_T$ is the Bond albedo of an emitting facet, $\xi_T$ is its local zenith angle, $S_T$ is its surface, $\zeta_{T}$ is the angle between the normal of the transmitter and the receiver facets, $\zeta_{R}$ is the angle between the normal of the receiving and the emitting facets, and $\delta_T$ is the distance between the two facets.
The infrared contribution can be written as:
\begin{equation}
E_{IR} = \sum_T~ \varepsilon\sigma T_T^4~ \frac{S_T}{\pi}~~\frac{\cos{\zeta_{T}} \cos{\zeta_{R}}}{\delta_T^2 } 
\label{e_ir}.
\end{equation}
In this equation, the surface temperature of each emitting facet is approximated using the energy balance: 
\begin{equation}
(1-\mathcal{A}_T) \frac{F_{\odot} \cos{\xi_T}}{r_H^2} = \varepsilon \sigma T_T^4,    
\end{equation}
which accounts for direct insolation only, without any prerequisite knowledge of the importance of the self-heating contributions. When an emitting facet experiences night during a given time step, we set a minimum threshold of $T_T$~=~20~K (various values have been tested and they do not result in any significant variation of the outcomes).

\subsubsection{Time step and orbital evolution}

Geometric calculations are performed with a cadence of one output every 8~minutes, over a full orbital revolution of the comet (i.e., $\sim$6.44~years). As a consequence, the thermal evolution model is run with a time step of 8~minutes, for ten full orbital revolutions. These ten revolutions represent the approximate time 67P has been evolving on its current orbit \citep{maquet2015}.
However, before running the thermal evolution simulations for ten revolutions on the current orbit of 67P, we simulate an injection of the nucleus from the Kuiper Belt to the inner solar system. We use a standard multistage injection process, described by several successive orbits with semimajor axis and eccentricity values given in Table \ref{multistage}. These allow for the slow regression of ice sublimation fronts, and the amorphous-crystalline water ice boundary, below the surface, which mimic the thermal processing sustained by 67P prior to its current orbit \citep[see, for example,][]{gkotsinas2022}.

\begin{table}[!h]
    \centering
    \caption{Parameters of the multistage injection orbits.}   
    \begin{tabular}{cccc}\hline
    Orbit & a [au] & e & q [au] \\
    \hline \hline
    Multistage 1 &50  &0.5  &25 \\
    Multistage 2 &25  &0.4  &25  \\
    Multistage 3 &8  &0.5  &4  \\
    \hline
    \end{tabular}
    \begin{tablenotes}
      \small
       \item[*] {a: semimajor axis; e: eccentricity; q: perihelion distance.}
    \end{tablenotes}
    \label{multistage}
\end{table}


    \begin{figure*}[h!]
        \centering
        \includegraphics[width=\textwidth]{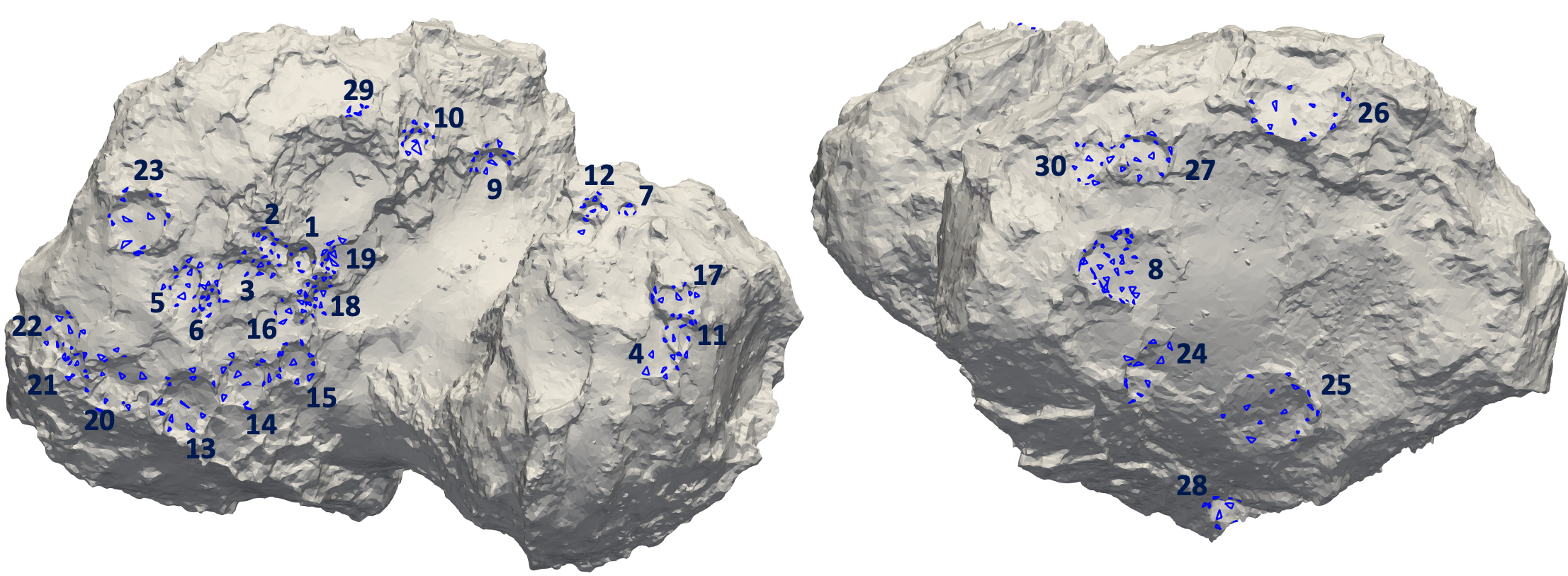} 
        \includegraphics[width=\textwidth]{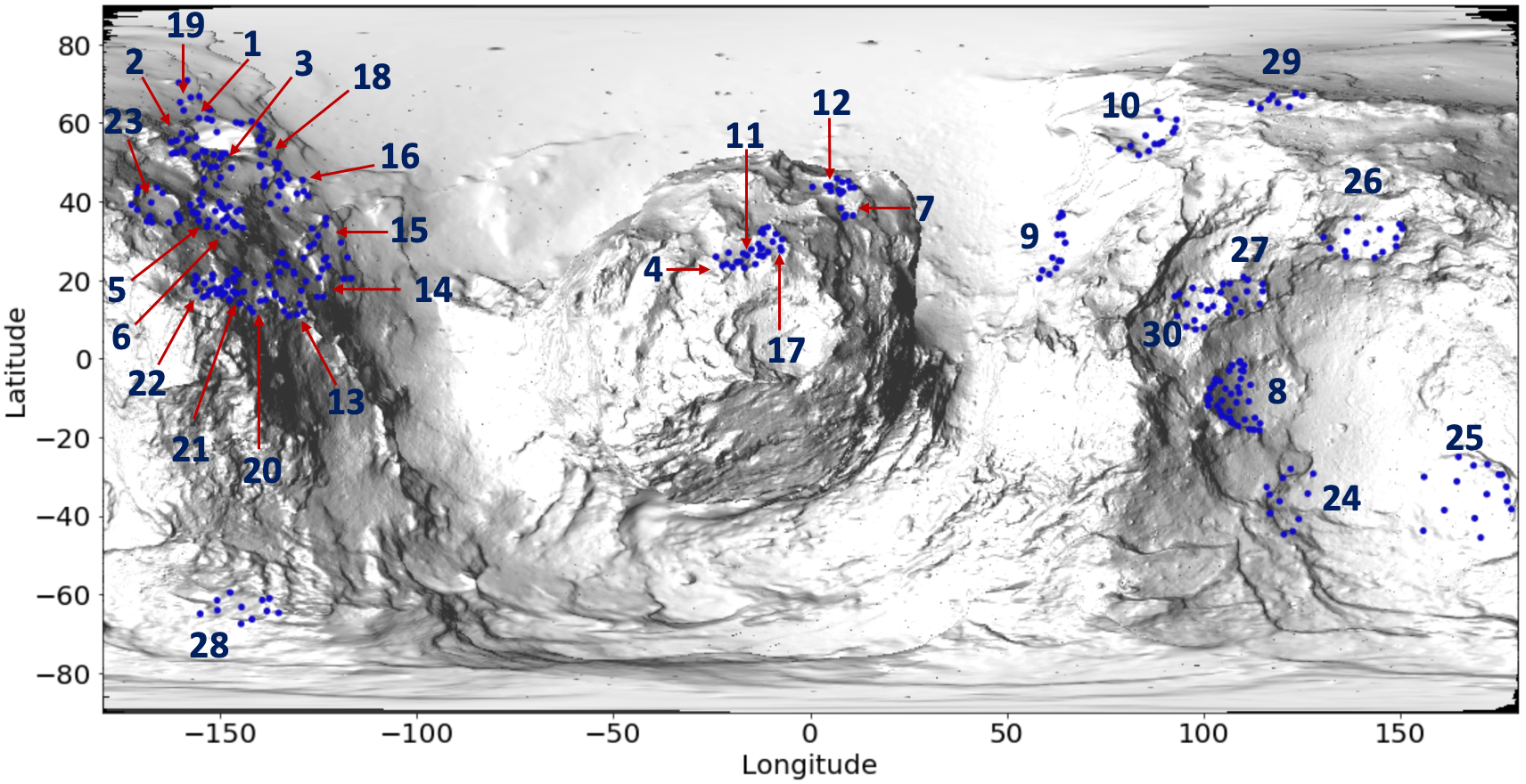} 
        \caption{Display of the facets selected for the study of the 30 pits. Top: the location of the facets on the surface of 67P. The shape model presented is the SPG model composed of 124,938 facets \citep{preusker2017}, which is used for the surface energy calculation. Bottom: the location of the facets on a 2D map of 67P, which is a projection of the high-resolution SPG shape model composed of 12~million facets.}
        \label{location3d}
    \end{figure*}

\subsection{Selection of pits on the nucleus} 
The diversity of local morphological features at the surface of 67P has been recently reviewed by \citet{el-maarry2019}. Most circular depressions can be found in the northern hemisphere, where deep pits and steep cliffs are also observed. Pits in the southern hemisphere are scarcer, and typically wider and shallower than the ones found in the northern hemisphere. This dichotomy is explained by strong seasons affecting the nucleus, where the southern hemisphere sustains intense heating and erosion during the summer \citep{keller2015}.
For this study, we selected pits with different shapes and dimensions that can be representative of the different illumination conditions at the surface, on both hemispheres and on both lobes, with as much sampling in latitude as possible. We focus on large features, and therefore do not include cometary thermokarst depressions, for example \citep{bouquety2021}.

With these constraints in mind, we selected 30 pits: their positions on the surface, as well as morphological characteristics such as their approximate diameter and depth, are given in Table~\ref{table_pos}. We note that not all features are circular or quasi-circular pits. Indeed, we also selected elongated pits and alcoves, as well as cliffs, in order to achieve our sampling goals and study possible evolutionary links between those features. 
We further note that some of the pits selected have shown activity, witnessed by \emph{Rosetta}/OSIRIS \citep{vincent2015a}.
For each pit, facets of the shape model were selected on the surrounding plateaus, the bottom, and the walls (see example in Fig.~\ref{pit5}, panel A) for a detailed study of each energy contribution (direct insolation, self-heating, and shadowing), and thermal evolution.
One caveat of our method is that we do not account for any shape evolution. Indeed, it is impossible to know what these morphological structures looked like ten orbits ago, as we still do not know how, when, and through what process they were formed. Therefore, the erosion sustained at each time step is not used to modify the geometry of morphological structures. Instead, erosion after ten cometary orbits is assessed from the current shape of 67P's nucleus, as observed by \emph{Rosetta}.

All facets selected for our study can be localized on the 3D shape model (Fig.~\ref{location3d} top panel) and a 2D map of the high-resolution shape model in an equidistant cylindrical projection (Fig.~\ref{location3d} bottom panel). We note that this map is based on a 12~million facet version of the SHAP7 shape model: we created a ``rubber sheet'' by putting each vertex point at an elevation proportional to its distance from the comet center above the plane of evenly spaced latitude and longitude. Shading was then realized through 3D rendering. The equidistant cylindrical projection cannot display the overhung areas, but we do not study any such feature here. We note that sophisticated map projections that do display the complete surface of the comet have been presented \citep[e.g.,][]{grieger2019,leon-dasi2021}.


\section{Case of one pit: Assessing the influence of initial parameters} \label{initial-params}

\begin{figure}[!h]
        \centering
        \includegraphics[width=\columnwidth]{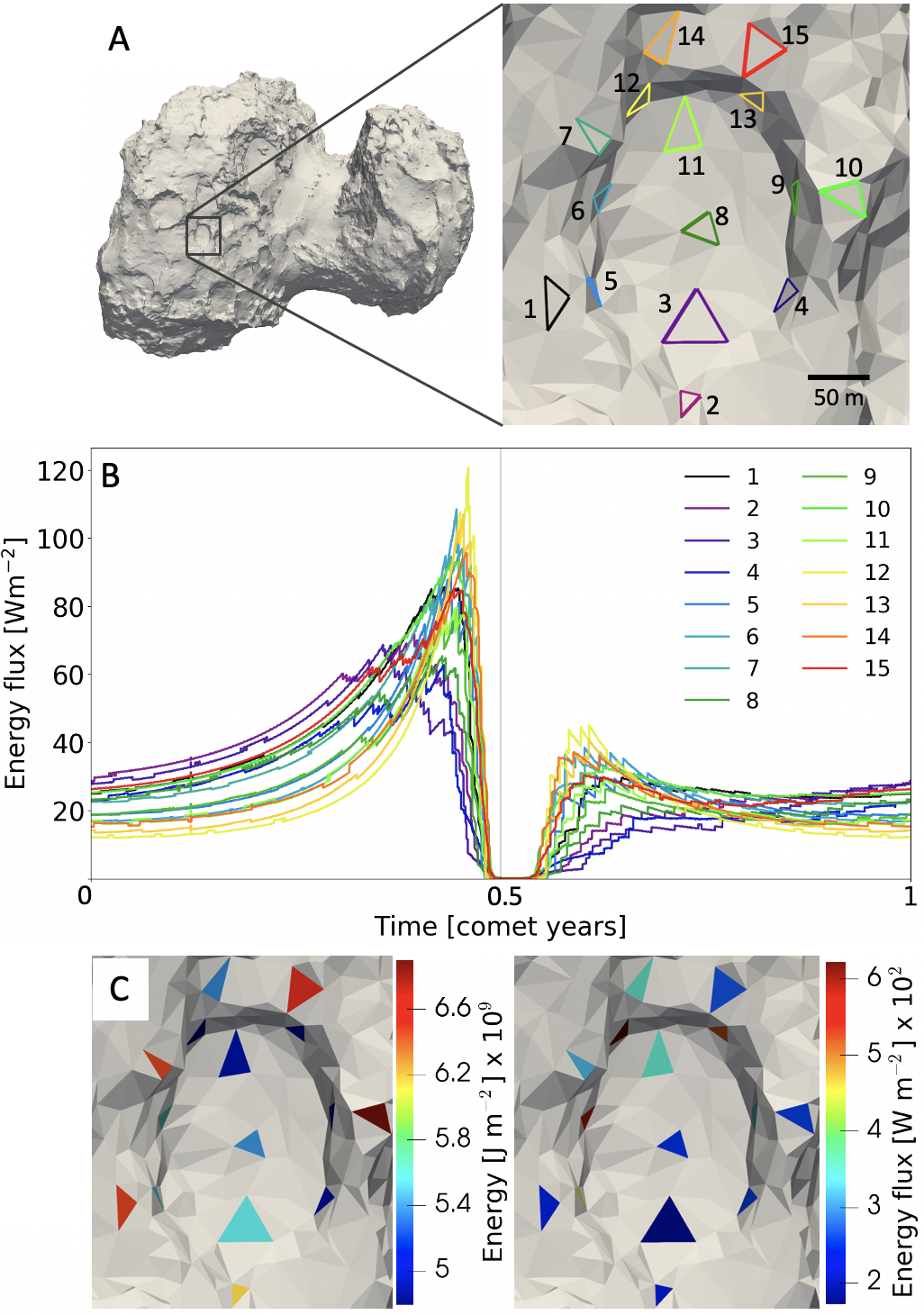} 
        \caption{Pit selected for the study of the influence of initial parameters, and corresponding energy. A: the location of pit 5, and facets sampled on the plateaus, walls, and bottom. B: the energy received at the 15 facets over one complete orbit, averaged over a daily period window. The gray line marks the perihelion. C: the total quantity of energy integrated over one orbit (left) and the maximum reached during the perihelion (right).}
        \label{pit5}
\end{figure}

In this section, we study the case of one pit to understand the influence of each critical initial parameter on its evolution.
The effect of these parameters on the outcomes of our thermal-evolution model will need to be kept in mind when discussing our results.
To avoid any interference between parameters, each of them is studied independently of the others. This pit is located in the Seth region and highlighted in panel A of Fig. \ref{pit5} (label 5 in Table \ref{table:position_coord}). It is on the big lobe's northern hemisphere, away from the influence of shadowing by the small lobe, so as to avoid self-heating contributions due to the global shape of the nucleus. Hence, only self-heating due to the local topography of the depression can influence its evolution. Fifteen facets were selected at the bottom, on the walls, and on the plateau surrounding this pit.
The energy received on the surface, which includes shadowing and self-heating contributions, is shown in panels B and C of Fig. \ref{pit5}.

\subsection{Influence of porosity} \label{sec:porosity}
\begin{figure}[!h]
    \centering
    \includegraphics[width=\columnwidth]{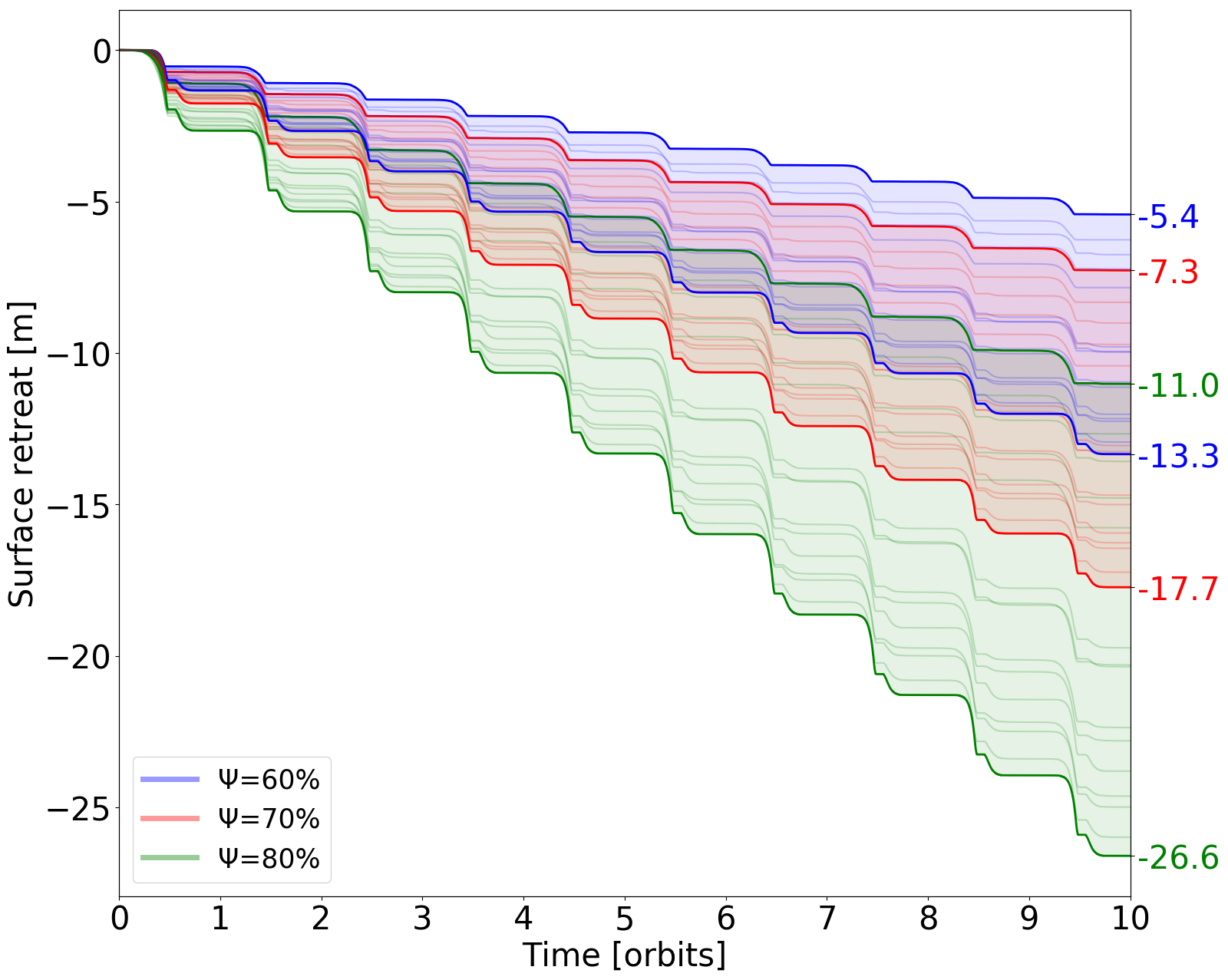}
    \caption{Progressive erosion sustained during ten full revolutions on 67P's current orbit, for all facets of pit 5, and three values of the porosity: 60\% (blue), 70\% (red), and 80\% (green). Vertical lines and numbers correspond to aphelion passages.}
    \label{psi_1d}
\end{figure}

\begin{figure*}[h!]
    \centering
    \includegraphics[width=\textwidth]{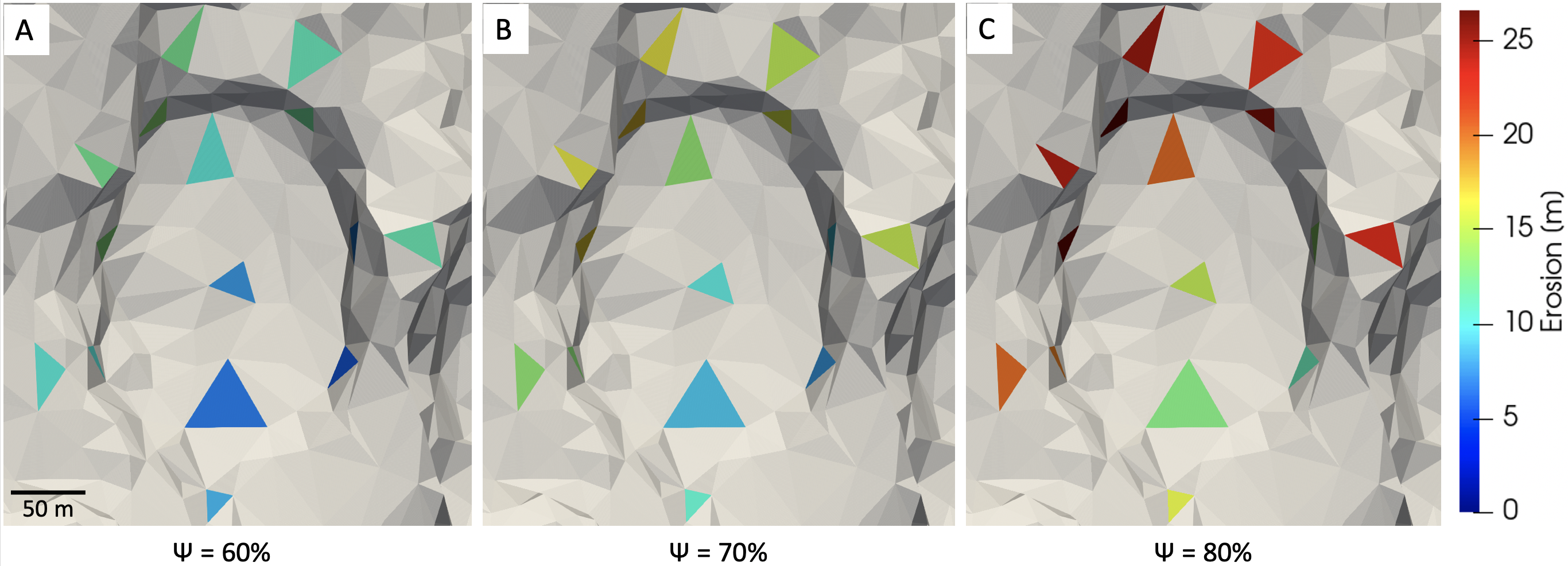} 
    \caption{Erosion sustained after ten orbital revolutions for each facet of pit 5, and different values of porosity: 60\% (A), 70\% (B), and 80\% (C).}
    \label{psi_3d}
\end{figure*}

The \emph{Rosetta} data allowed us to derive values for the internal porosity ranging from 75\% to 85\% \citep[][for example]{herique2016}. However, some areas of the surface appear to be consolidated material \citep{el-maarry2019}, with a likely lower local porosity, although direct measurements have not been made. 
We thus tested three values for this parameter, in order to assess its influence on the outcomes of our thermal evolution simulations: 60\%, 70\% and 80\%. From these, we see that a higher porosity, $\psi$, induces a larger amount of erosion. Indeed, the total mass of eroded material is essentially the same in the different tests, driven by the total amount of energy received locally by each facet. The volume of this corresponding mass varies, however, increasing with an increasing porosity. As such, the extent of the erosion sustained after ten revolutions for $\psi$ ~=~70\% is (on average for all facets) $\sim$30\% higher than with $\psi$~=~60\%, and the erosion for $\psi$~=~80\% is $\sim$50\% higher than for $\psi$~=~70\% (see Figs.~\ref{psi_1d} and \ref{psi_3d}).

\subsection{Influence of the dust-to-ice mass ratio} \label{sec:dusttoice}
\begin{figure}[!h]
    \includegraphics[width=\columnwidth]{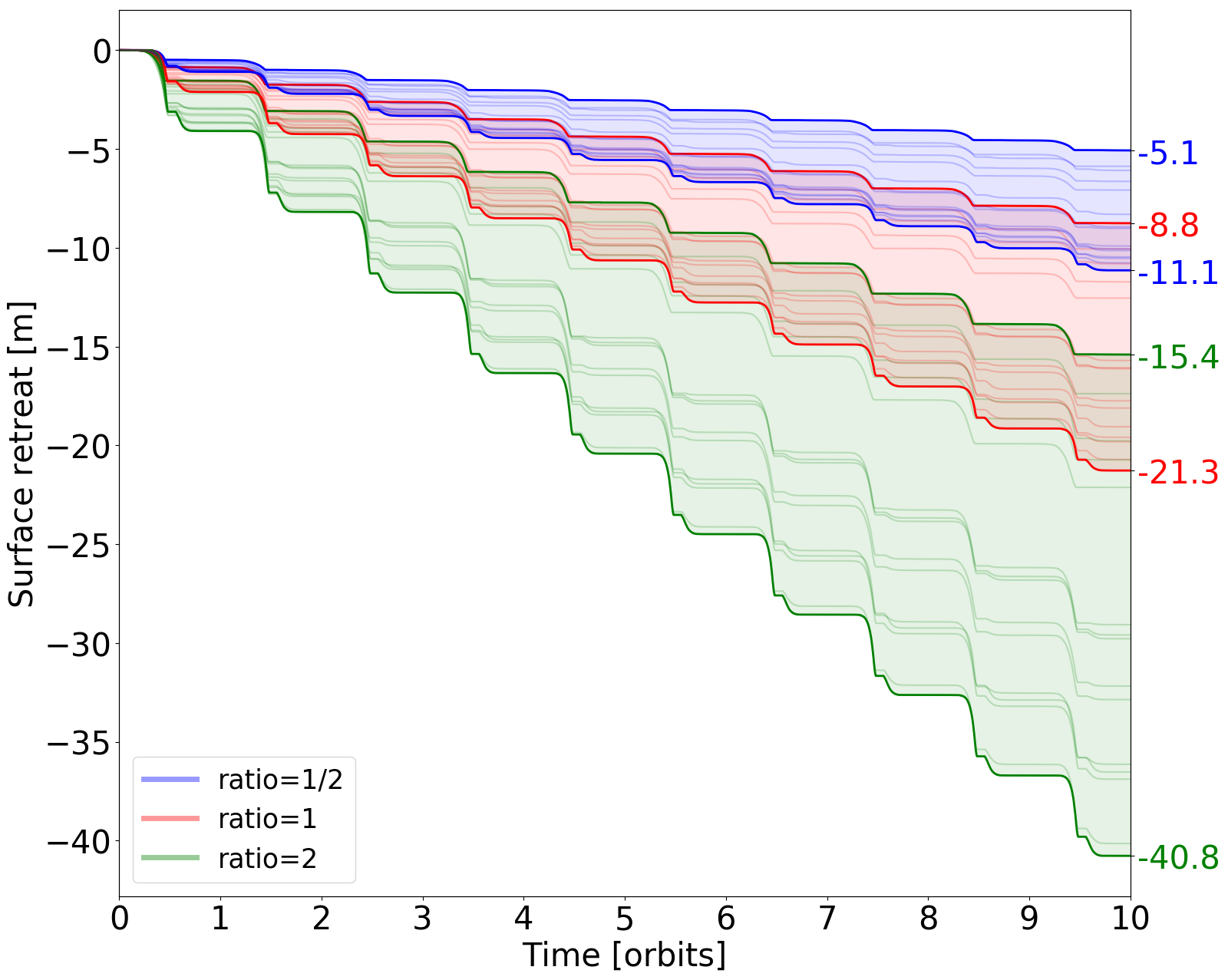}
    \caption{Progressive erosion sustained during ten full revolutions on 67P's current orbit, for all facets of pit 5, and three values of the dust-to-ice mass ratio: 0.5 (blue), one (red), and two (green). Vertical lines correspond to aphelion passages.}
    \label{ratio_1d}
\end{figure}

\begin{figure*}[!h]
    \centering
    \includegraphics[width=\textwidth]{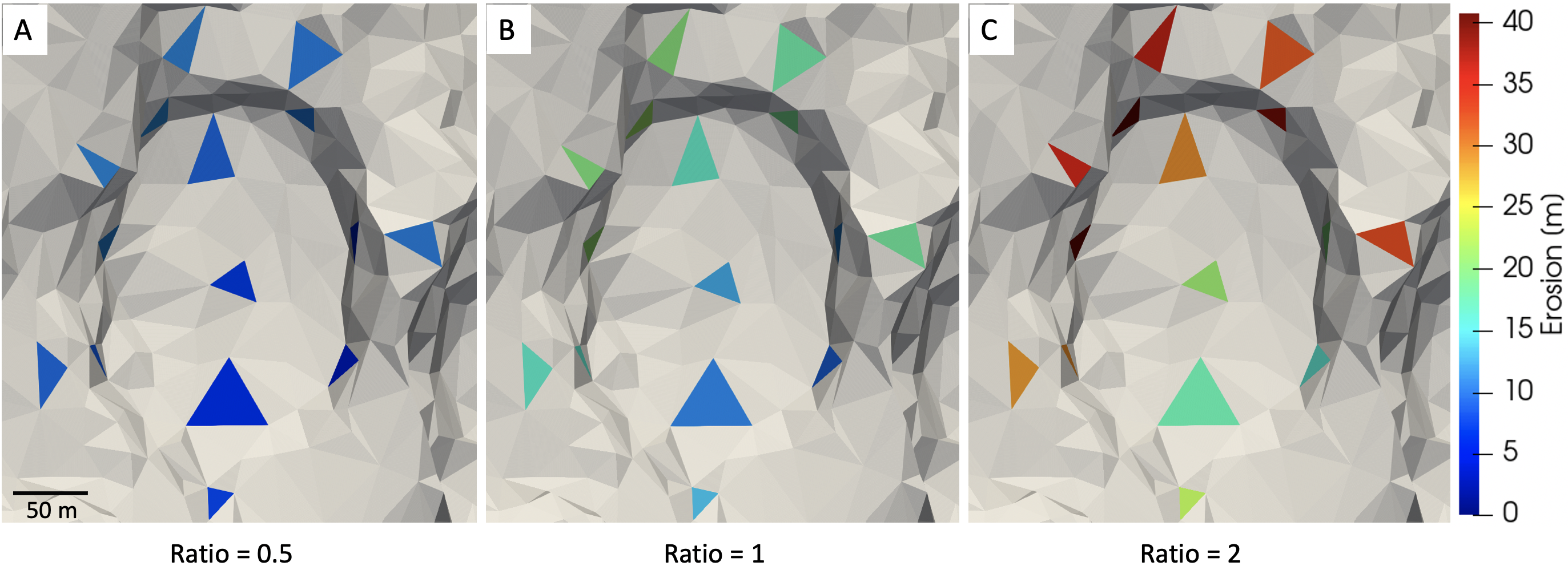} 
    \caption{Erosion sustained after ten orbital revolutions for each facet of pit 5, and different values of the dust-to-ice mass ratio: 0.5 (A), one (B), and two (C).}
    \label{ratio_3d}
\end{figure*}

The bulk dust-to-ice mass ratio of a cometary nucleus is notoriously difficult to constrain, especially if it is inferred from the coma composition \citep{choukroun2020}. For modeling the thermal evolution of comets, a value of one has historically been used, from the \emph{Giotto} mission measurements at 1P/Halley \citep[see][and references therein]{huebner2006}. It appears to be consistent with the \emph{Rosetta} measurements for 67P \citep{choukroun2020}.
As seen in Figs.~\ref{ratio_1d} and \ref{ratio_3d}, we tested the effects of different dust-to-ice mass ratios, 0.5, one, and two, to assess its influence on the thermal evolution.
As for porosity, the initial dust-to-ice mass ratio has a significant influence on the extent of the erosion sustained after ten full orbital revolutions. In fact, we see that erosion substantially increases with an increasing dust-to-ice mass ratio: it almost doubles when we double the ratio. This result is consistent with how the dust-to-ice mass ratio influences the value of thermophysical parameters, such as the thermal conductivity of the material. Indeed, as the dust-to-ice mass ratio increases, so does the thermal conductivity. Energy is thus transferred into deeper layers of the subsurface, leading to the sublimation of deep ice, and thus erosion at relatively higher depths.

\subsection{Influence of the CO and CO$_2$ abundance} \label{sec:abundance}

The abundance of CO and CO$_2$ in a cometary nucleus is also difficult to measure. Their bulk abundances are usually assessed from production rates measured in the coma, but the procedure is not straightforward. From modeling experiments \citep[e.g.,][]{prialnik2006}, we know that values integrated over a long period of time are more accurate than data obtained at a single moment on the orbit. \citet{herny2021b} showed that the nucleus of 67P can be considered uniform at the first order. Production rates of CO and CO$_2$ do vary significantly across the orbit, sometimes by several orders of magnitude \citep{fougere2016,biver2019,lauter2019,combi2020}.
Besides, these volatiles are very sensitive to cumulative heating; hence, the uppermost surface layers are certainly depleted compared to the bulk values.

\begin{figure*}[!h]
    \centering
    \includegraphics[width=\textwidth]{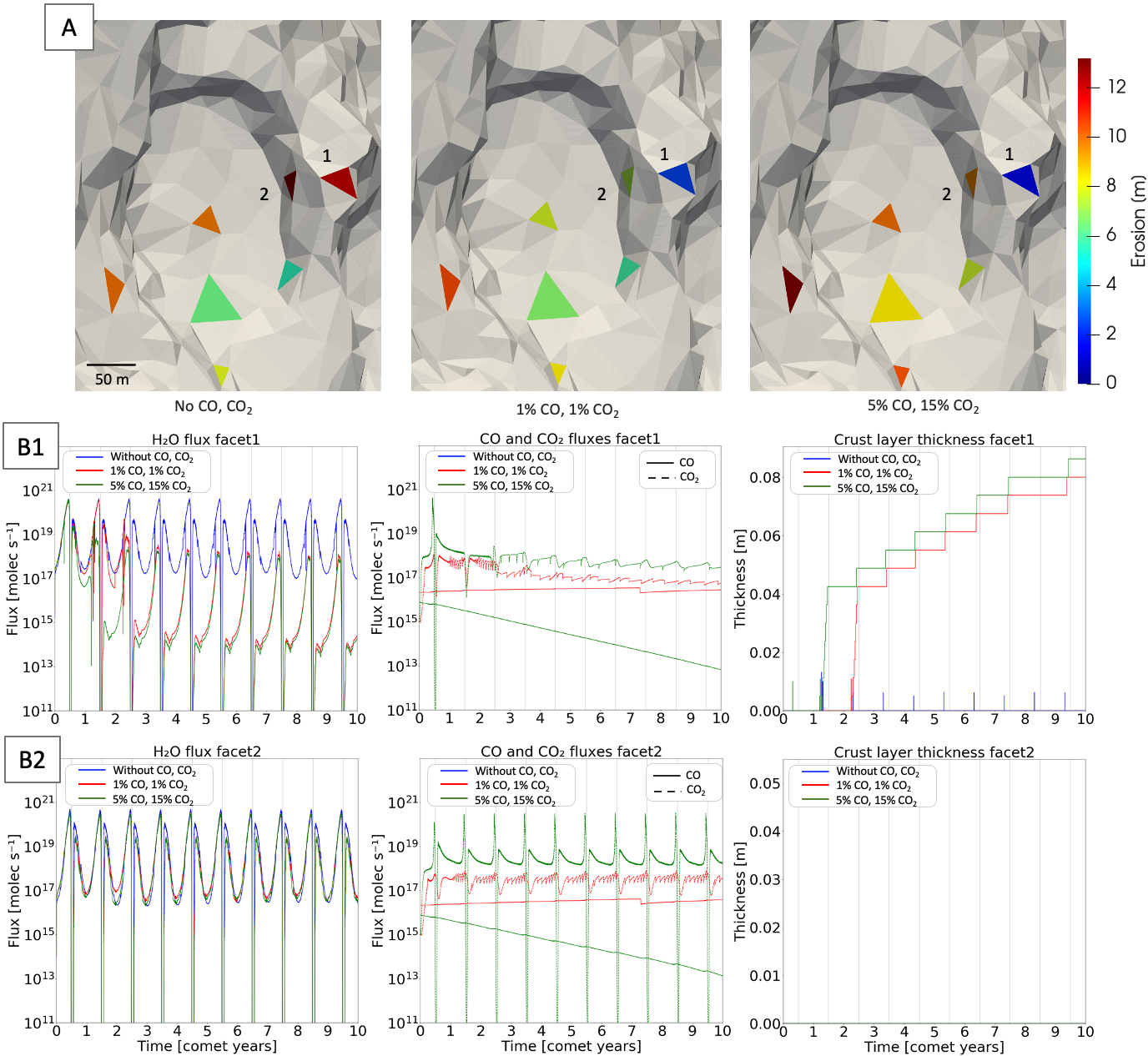}
    \caption{Influence of adding CO and CO$_2$ to the ice composition. A: the erosion sustained after ten orbital revolutions for seven facets of pit 5, for various CO and CO$_2$ abundances: no CO and CO$_2$, 1\% CO and 1\% CO$_2$, and 5\% CO and 15\% CO$_2$ from left to right, respectively. B1 and B2: the activity patterns of two facets are given -- H$_2$O production rate on the left, CO$_2$ and CO production rates in the middle column, and thickness of the dust layer on the right. The activity of facet 1 is quenched in the presence of CO and CO$_2$. The behavior of facet 2 is given for comparison: it remains active in the presence of CO and CO$_2$, preventing the formation of a dust layer (right plot).}
    \label{volatiles_3d_gflux}
\end{figure*}

We tested the effect of the presence of CO and CO$_2$ by setting their initial mass fraction with respect to water to various values: 0\% for both, 1\% for both, and 5\% CO with 15\% CO$_2$. 
Adding such large amounts of CO and CO$_2$ to the initial inventory of volatiles, even if their sublimation boundary regresses below the surface during the multistage injection process, triggers some numerical instability for many facets. 
Fixing this numerical instability requires that we change the initial thermophysical parameters for these facets, which would defeat our purpose. Therefore, we do not compare the unstable facets further in this analysis (for instance, this is why in Fig. \ref{volatiles_3d_gflux} we cannot display results for the 15 facets, as in the prior cases).
We do, however, detail in the discussion the potential origin and implication of these facets' numerical behavior.
For facets that do not suffer from numerical instability, the sublimation of water ice remains the main driver for both activity and erosion. 
Most importantly, we note that all facets do not behave with the same pattern, in a departure from what was observed in previous tests, where all facets would follow the same trend: this effect can be observed in panel A of Fig.~\ref{volatiles_3d_gflux}. Facets that receive the largest amount of energy integrated over one orbit are very active. However, they tend to build a dust layer at the surface after several perihelion passages, which prevents them from being active during subsequent perihelion passages (see panel B1 of Fig.~\ref{volatiles_3d_gflux}). Facets receiving lower amounts of energy, on the other hand, are not active enough to build up such a dust layer. They do get a layer of dust at the surface, but of insufficient thickness to completely quench subsequent activity. Hence, they remain active for the ten full orbital revolutions (see panel B2 of Fig.~\ref{volatiles_3d_gflux}). The dust production rate is, overall, similar for all facets remaining active throughout the ten orbital revolutions, whatever the amount of CO and CO$_2$ present in the icy phase. Facets for which the activity is quenched, however, stop emitting dust in the coma.

Overall, adding CO and CO$_2$ to the initial composition leads to a substantial complexity in the thermal simulations (numerical instabilities and unpredictable behavior), without significantly altering the outcomes when we simulate the thermal evolution for ten full orbital revolutions.

\subsection{Thickness of the surface dust mantle} \label{sec:mantle}
As highlighted by the results obtained in the previous tests, the thickness of the dust mantle at the surface may play a key role in the evolution of morphological features such as pits. The long-term survey of 67P's nucleus by \emph{Rosetta} has revealed that large smooth plains in the northern hemisphere are covered by a dust mantle, which originates from the southern hemisphere \citep{thomas2015}, following the ejection of dust particles around perihelion \citep{keller2015, keller2017}. The thickness of this mantle is unknown and almost certainly nonuniform \citep{hu2017, davidsson2021}. However, \citet{davidsson2022} suggest that the dust mantle in the northern hemisphere may typically be thinner than 2~cm. 
Interestingly, \citet{herny2021b} noted that the assumption of the presence of a thin dust mantle was required in order to fit the Rosetta Orbiter Spectrometer for Ion and Neutral Analysis/Double Focusing Mass Spectrometer (ROSINA/DFMS) measurements, in particular to reproduce the patterns of volatile production rates. We have seen from the simulation outcomes discussed previously that a thin dust mantle naturally forms at the surface, without affecting the general results in terms of activity and erosion. Indeed, such a thin dust mantle is periodically produced and removed from the surface, after dust particles are dragged by escaping gas. 

\begin{figure}[!h]
    \includegraphics[width=\columnwidth]{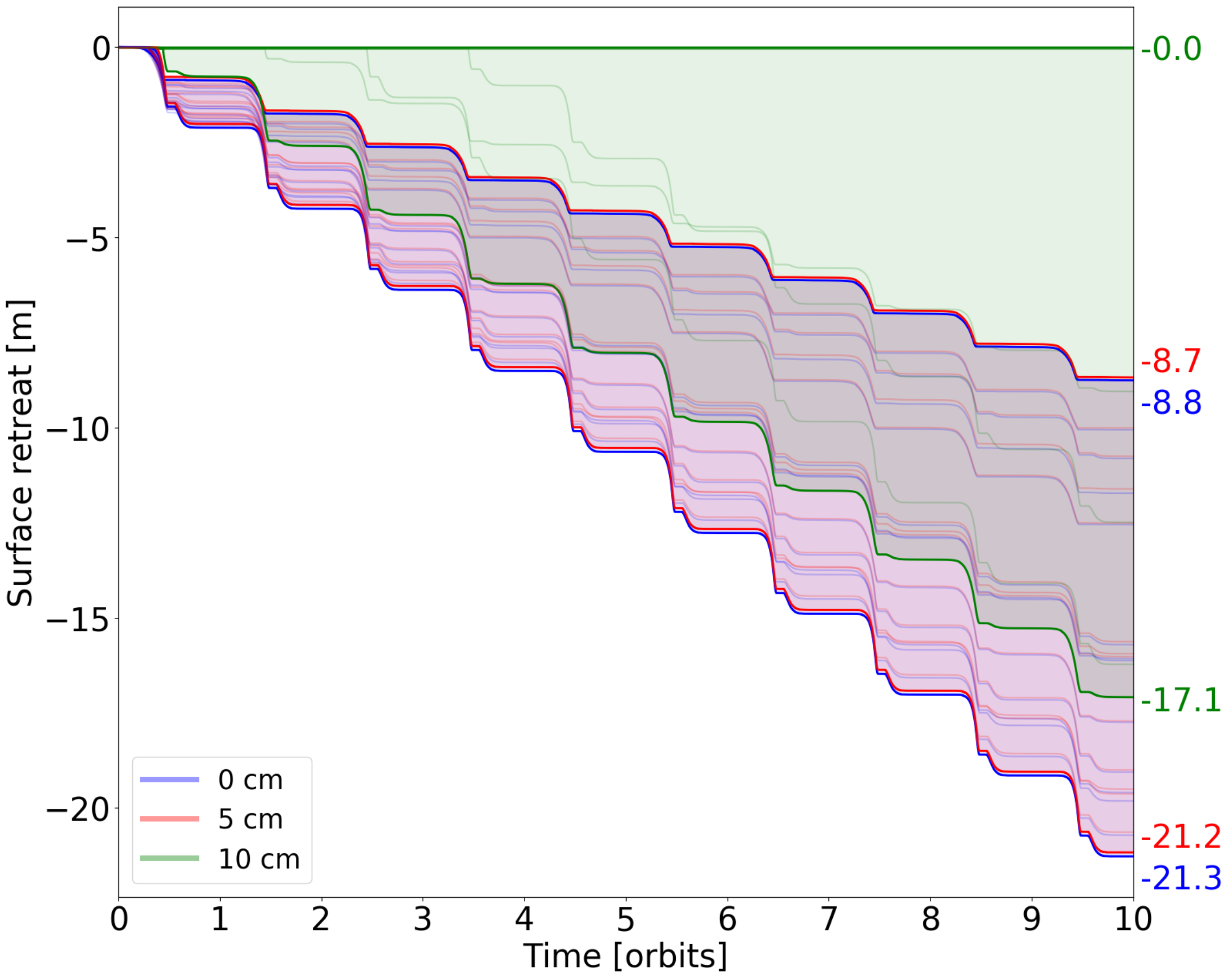}
    \caption{Progressive erosion sustained during ten full revolutions on 67P's current orbit, for all facets of pit 5, and three values of initial dust mantle's thickness: 0~cm (blue), 5~cm (red), and 10~cm (green). Vertical lines correspond to aphelion passages.}
    \label{mantle}
\end{figure}

\begin{figure*}[h!]
    \centering
    \includegraphics[width=\textwidth]{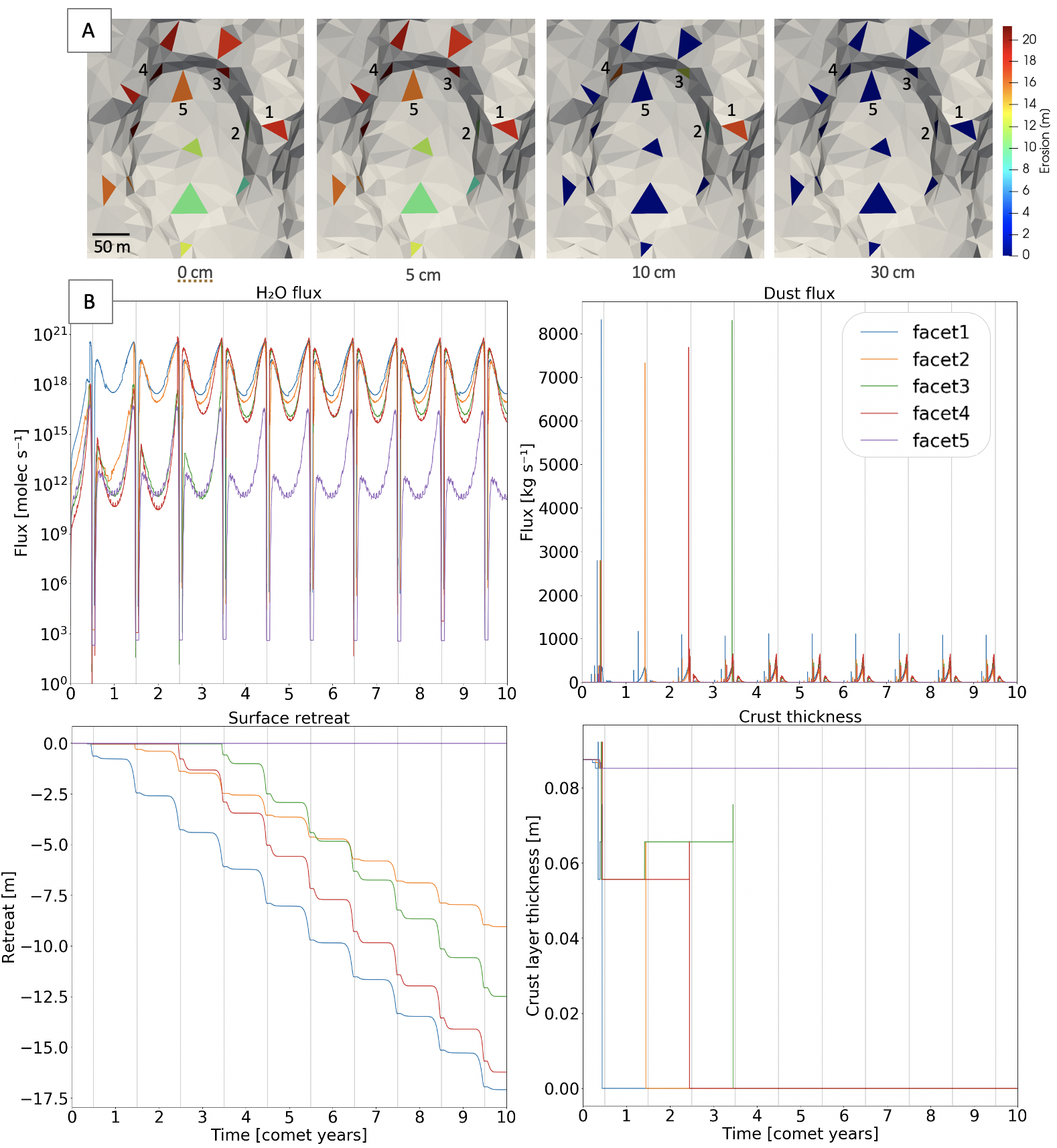} 
    \caption{Influence of the presence of a dust mantle. A: the erosion sustained after ten revolutions with an initial dust mantle of 0~cm, 5~cm, 10~cm and 30~cm, left to right, respectively. B: the H$_2$O production rate, the dust production, the progressive erosion, and the thickness of the dust layer, given for five facets, for an initial dust mantle of 10~cm.}
    \label{mantle_3d_10cm_4outputs}
\end{figure*}

We further tested the influence of the thickness of a dust mantle, assuming that such a layer is initially present at the surface of the nucleus when the nucleus reaches its current orbit back in 1959 \citep{maquet2015}.  Several values for the thickness were tested: 5~cm, 10~cm, 30~cm, 60~cm, and 1~m. We find that a 5cm-thick dust mantle is easily removed by cometary activity after the first perihelion passage. In previous tests, such very thin mantles did form and were removed, as described in the previous section, whenever the appropriate conditions were met. The presence of thin dust mantles on the surface of the facets is thus expected from the previous simulation results. As a consequence, adding an initial dust mantle of 5~cm gives the same simulation outcomes as if no dust mantle was considered. 
When the simulations start with a 10cm-thick dust mantle, cometary activity is quenched for most facets of this pit (located in the northern hemisphere), as shown in Figs.~\ref{mantle} and \ref{mantle_3d_10cm_4outputs}. Only four facets remain active throughout the ten orbital revolutions. Their sustained activity might be attributed to two factors: they receive the most energy close to perihelion, which leads to a short period of strong activity that removes the dust mantle, or they receive the most energy integrated over one orbit, which allows them to eventually remove the dust mantle after several perihelion passages (Fig.~\ref{mantle_3d_10cm_4outputs} bottom panel). 
When the dust mantle thickness is larger than 10~cm (i.e., 30~cm, 60~cm, or 1~m), the activity remains quenched, and we do not observe any facets able to remove this layer during the ten orbital revolutions. We thus only show results for a thickness of 30~cm for comparison, in Fig.~\ref{mantle_3d_10cm_4outputs}.

\subsection{Set of uniform initial parameters} \label{sec:params_selected}

Whether cometary nuclei are homogeneous in composition or thermal and mechanical characteristics remains a matter of debate. Heterogeneities can be found at various spatial scales and might also be the result of evolution, due to initial nonuniform thermophysical properties, or simply a nonuniform insolation of the surface, due to the shape of the nucleus and seasonal variations \citep{guilbert-lepoutre2011}. Therefore, it is entirely possible that different processes, related to nonuniform characteristics, are at the origin of the formation and evolution of sharp morphological features such as pits. However, in this study, we seek to quantify how energy received at the surface -- through direct insolation, local and global self-heating, and shadowing effects -- may give rise to morphological features of the spatial scale observed by \emph{Rosetta}. 
Therefore, we seek a set of initial values for the thermophysical parameters, such that the same set can be used for all facets, on all our morphological features of interest, regardless of their location on the nucleus. These might not be representative of all the local conditions for all features studied, but are a good enough approximation to quantify local and global trends in the erosion rate. 
In the remainder of our study, we therefore consider a bulk porosity of 75\%, and a dust-to-ice mass ratio of one. No CO or CO$_2$ was included in the ice mixture, to avoid numerical instabilities, but also because our tests show that they do not contribute to any significant change in the resulting erosion patterns, after ten orbital revolutions. Finally, no initial dust mantle at the surface was added. However, we stress that the formation of such a mantle is a natural consequence of cometary activity: the cyclic formation and destruction of such a layer of dust deposit is fully taken into account in our thermal evolution model.
The previous sections detailed the effect of each of these critical parameters on the evolution outcomes for one specific pit. The following section will make use of the uniform set of parameters, to study 30 morphological features (circular and elongated pits, and alcoves) located across the surface of 67P.


\section{Evolution of 30 morphological features across the surface of 67P} \label{sec:thermal-simu}

\subsection{Energy received at the surface: General trends}

\begin{figure*}[!h]
        \centering
        \includegraphics[width=\textwidth]{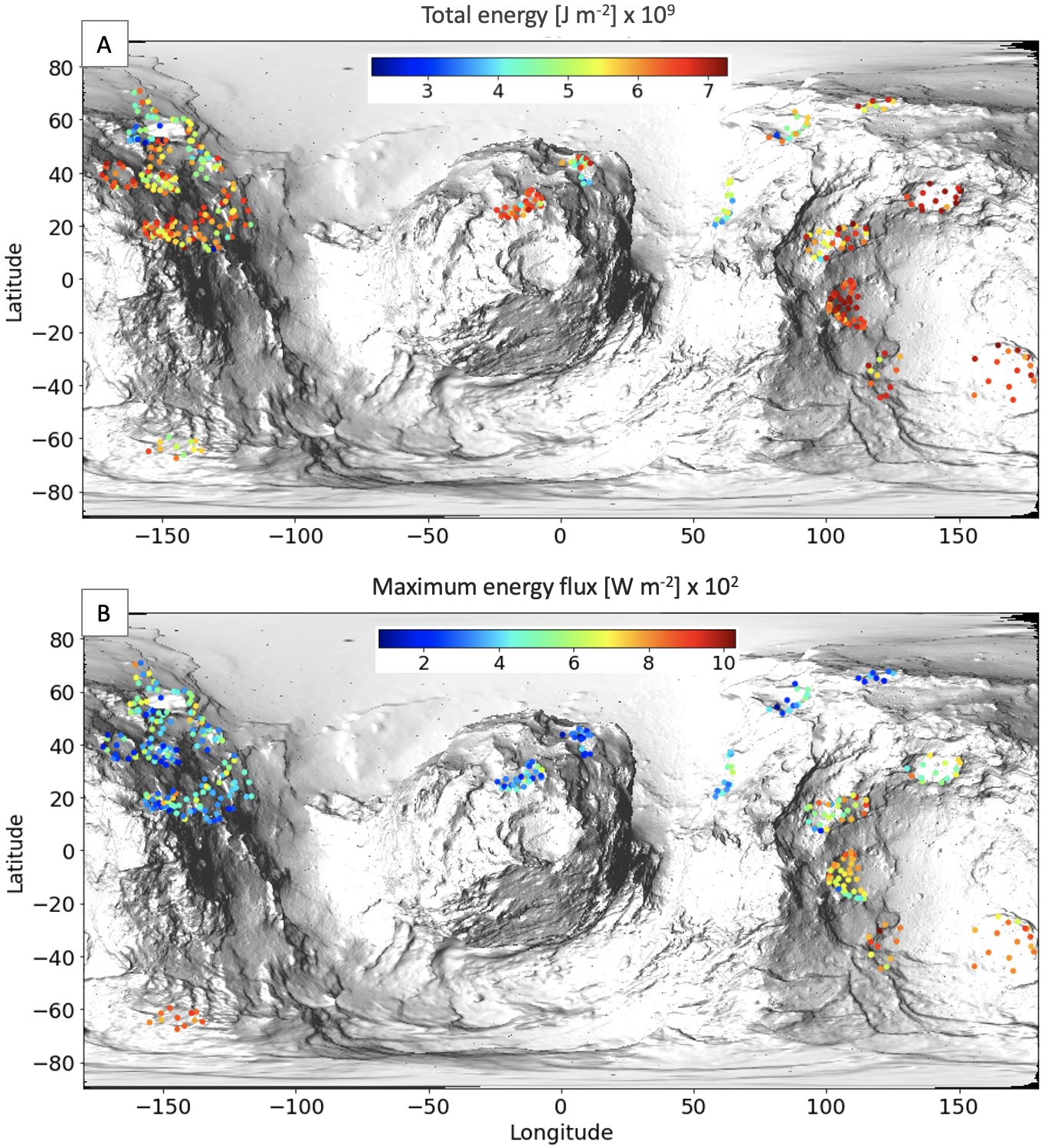}
        \caption{Energy flux received at the surface for all 380 facets, distributed over the 30 studied pits. A: the total quantity integrated over one complete orbit of 67P; B: the maximum reached during the perihelion passage.}
        \label{e_2D}
    \end{figure*}

\begin{figure}
    \centering
    \includegraphics[width=\columnwidth]{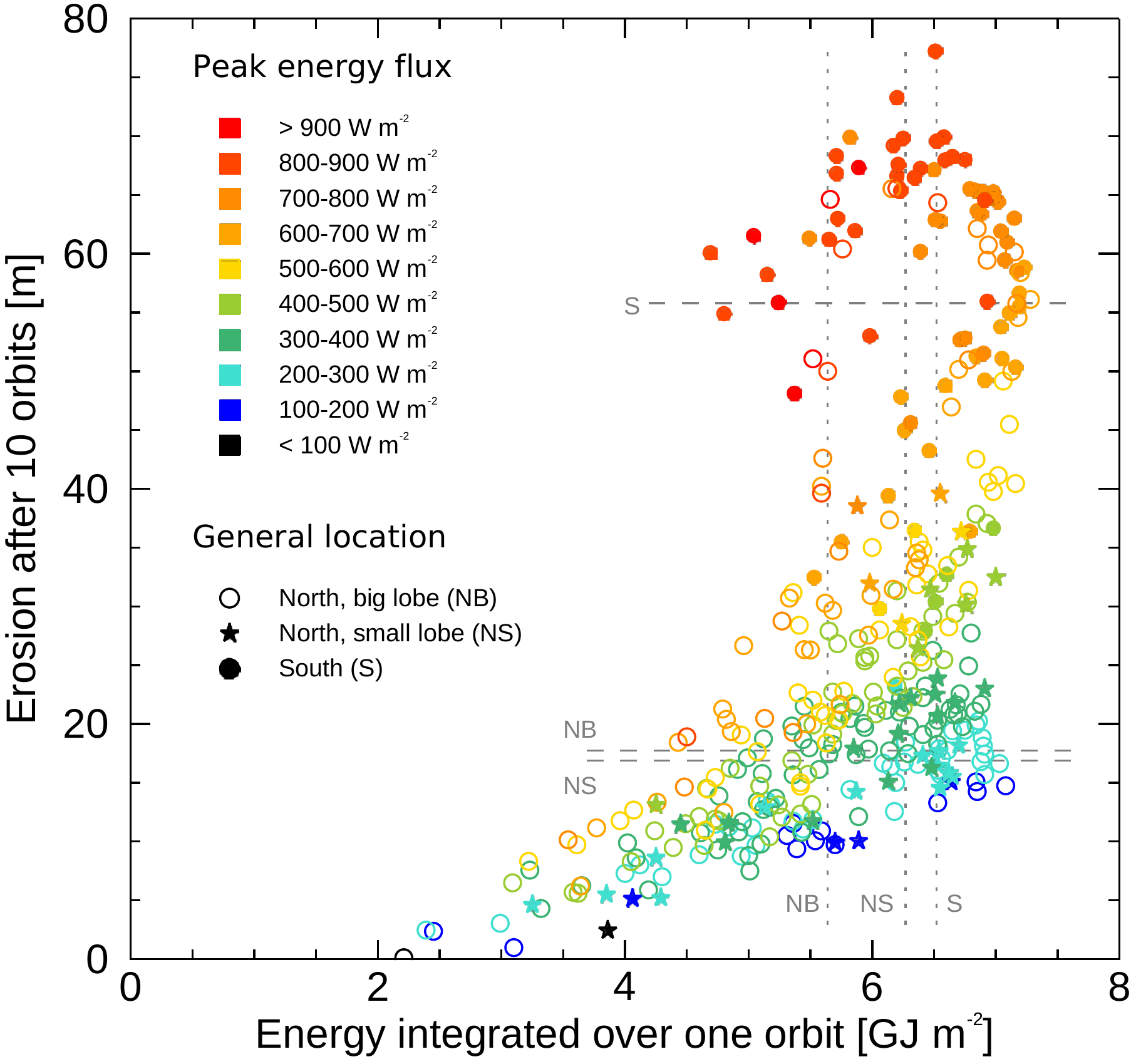}
    \caption{Erosion sustained at each facet as a function of the energy they receive, integrated over one orbit. The dotted gray lines show the median of this energy for the large regions, i.e., the small and big lobes on the northern hemisphere and the southern hemisphere. The dashed gray lines show the median of erosion sustained by the facets in these regions. The color code provides an indication of the peak energy received at or close to perihelion.}
    \label{Fig:nrjstat}
\end{figure}

To study the thermal processing of the 30 morphological features, a total of 380 facets were selected across the surface of 67P. For each facet, the energy input was computed, taking into account the effects of shadowing and self-heating as described in Sect. \ref{sec:methods}, and applied in Sect. 3. In Fig. \ref{e_2D} we show two quantities related to the energy input at the surface of 67P: the total energy integrated over one orbit, and the maximum energy flux received by each of the 380 facets. The maximum energy input is typically reached at perihelion for facets in the south, or just before perihelion for facets in the north.
These two quantities were found to be essential to interpret the results of the thermal evolution model.
Indeed, we see that the greatest amount of thermal processing -- inducing substantial water ice sublimation and erosion --  occurs during the perihelion passage, when the nucleus receives most of the direct solar energy on the southern hemisphere. As a result, the maximum energy is representative of this seasonal activity trend, and the maximum energy map does show the expected north-south dichotomy. However, we see from Fig.\ref{Fig:nrjstat} that a similar final amount of erosion can be achieved either by having a high amount of energy at or close to perihelion, or by having an increased amount of energy integrated over one orbit. 
As a result, we find facets in the north that can display some significant activity, out of the perihelion passage. This is due to the relatively high obliquity of the nucleus  \citep[$\sim$~52$\degree$,][]{sierks2015}. For those facets, considering the amount of energy received over the whole orbit is usually necessary to describe their thermal behavior. This is reflected in the lack of a clear north-south dichotomy in the integrated-energy map (see Fig.~\ref{e_2D}), as the northern hemisphere receives direct solar energy outside of the perihelion passage. 
However, we also see that for the same value of the integrated energy, facets that receive the bulk of the energy at perihelion tend to erode more (Fig.~\ref{Fig:nrjstat}).
Finally, these trends are not an absolute rule, and this justifies using a full subsurface thermophysical model, as we do below, rather than simply assessing the thermal behavior from surface energy balance maps. 

\subsection{Effects of local topography and global shape} \label{sec:shape_effects}

To the first order, the distribution of energy (integrated or maximum) is dominated by expected seasonal effects. We additionally see some variations for a given latitude and amongst facets related to one morphological feature. These can be generally attributed to shadowing and self-heating effects. 
Shadows are cast at the surface of 67P on a large scale (e.g., the neck area between the two lobes), as well as on a small, topographic scale (e.g., the bottom or part of the walls of deep circular pits). These can induce a significant decrease in the energy input, by as much as 70\%, depending on the facets' location and orientation. The effect of shadows can, however, be slightly offset by self-heating from neighboring facets (Fig.~\ref{sh_3d}).
For most pits, self-heating contributes less than 20\% of the total energy received at the surface. Thus, direct insolation dominates the energy input and self-heating is not the main activity driver. 
However, for several complex topographic configurations, where facets are not easily reached by direct insolation, self-heating can exceed the contribution from direct insolation. For these specific facets, the contribution of self-heating can reach more than 60\% of the total energy received at the surface (Fig. \ref{sh_3d}). They are typically located on the walls and at the bottom of deep circular pits. On a larger spatial scale, we also find such facets on alcoves close to the neck region, which are periodically in the shadow of the small lobe, and thus receive self-heating from it. 

\begin{figure*}[ht]
        \centering
        \includegraphics[width=\textwidth]{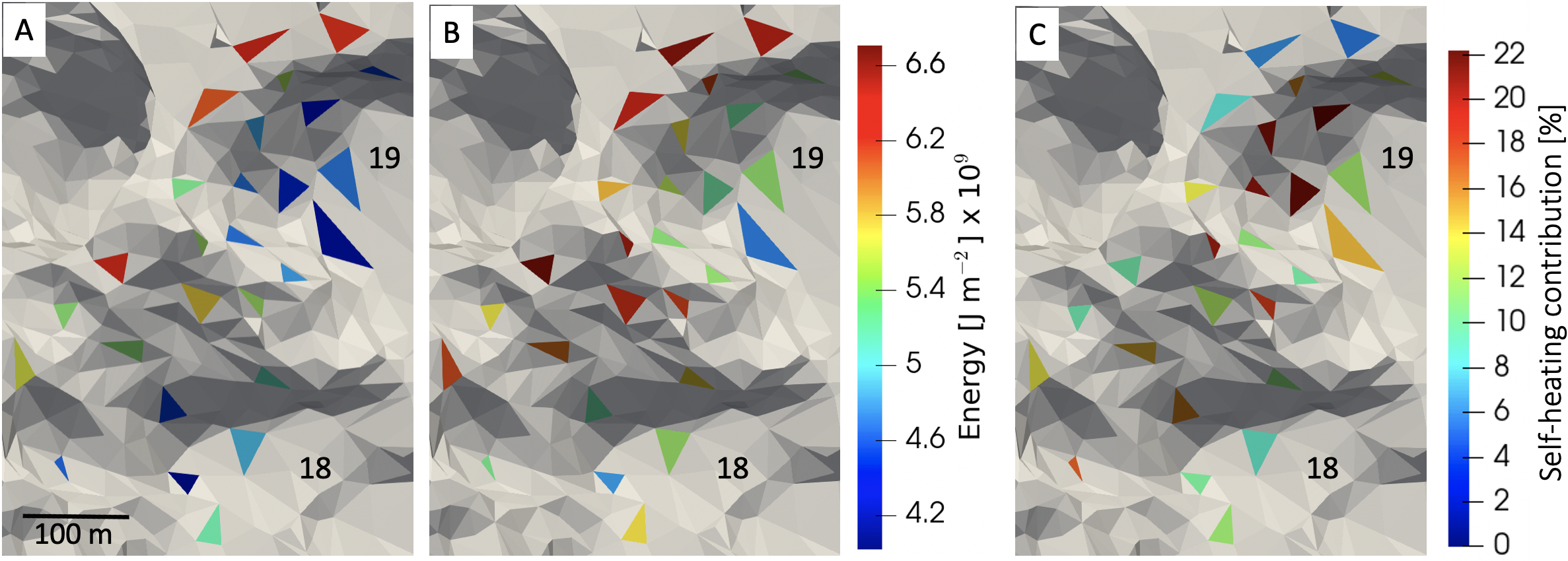} 
        \caption{Effects of the small lobe on the facing cliffs. A and B: the energy received at the surface of alcoves 18 and 19 integrated over one complete orbit with and without the small lobe in the shape model, respectively. C: the contribution of self-heating received from the small lobe only to the total energy received at the surface of structures 18 and 19.}
        \label{e_nolobe}  
    \end{figure*}

For the sake of completeness, we seek to quantify the relative contributions of the local topography versus the global morphology of the nucleus to the amount of self-heating.
We thus compare the energy input for some facets of the shape model, and the energy input of the same facets when we numerically remove the small lobe from the shape model. This comparison is most informative for features 18 and 19 (also known as Seth\_05 and Seth\_04, respectively). These are two alcoves located close to the neck area, whose evolution is extremely affected by the presence of the small lobe.
The integrated energy received over one orbit, with and without the small lobe in the shape model, is given in Fig.~\ref{e_nolobe} (panels A and B). Facets on the alcoves receive up to 70\% more energy when the small lobe is absent, due to the direct insolation reaching them. A detailed look at the various energy contributions informs us that the decrease in energy input from self-heating is not as significant as expected.
For facets located at the bottom of those alcoves, direct insolation becomes the dominant source of energy, as expected, although the contribution of self-heating does not drop to zero. For one facet, there is even a slight increase in the self-heating contribution ($\sim$7\%). 
This is due to the fact that surrounding facets receive much more direct insolation, and hence can transmit more energy. 
Overall, the contribution of the small lobe (vs. local topography) to the input of self-heating is not dominant (see Fig.~\ref{e_nolobe}). The small lobe contributes to up to $\sim$22\% of the total energy received by features 18 and 19. This is almost half of the total self-heating contribution for these alcoves, located in a region of the nucleus where the global contribution is maximum. However, in other regions, local topography is the major source of self-heating. 

\subsection{Thermal evolution simulations} \label{sec:erosion}

\begin{figure*}[ht]
        \centering
        \includegraphics[width=\textwidth]{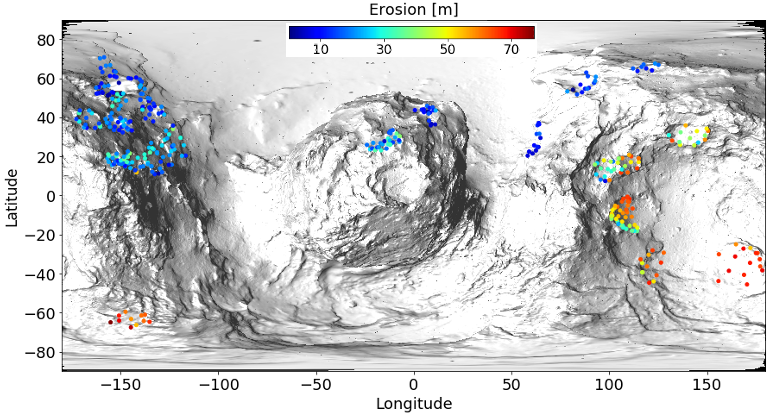} 
        \caption{Erosion sustained after ten revolutions on 67P's current orbit for all the 380 facets studied.}
        \label{er_2d}
    \end{figure*}

The energy received at the surface of each facet, with the global distribution and trends described above, is the boundary condition for thermal evolution simulations performed over ten full orbital revolutions. This energy input was used to quantify the activity for each facet, for example, phase transition, gas production, dust mantling, and erosion. To keep our model relatively simple, we did two things: 1) we used a uniform set of initial parameters for each facet, as derived from Sect. \ref{initial-params}; and 2) we did not account for any influence of shape evolution on the illumination conditions (i.e., erosion sustained at each time step was not used to modify the geometry of morphological structures). Instead, erosion after ten cometary orbits was assessed from the current shape of 67P's nucleus, as observed by \emph{Rosetta}. 
Global results, obtained for the 380 facets across 30 morphological features on the surface of 67P are represented in Fig.~\ref{er_2d}. 

\subsubsection{Latitudinal variations}

We see that erosion at the surface is mostly correlated with the energy received at or close to perihelion. As a result, a stark contrast is observed between both northern and southern hemispheres. 
After ten orbital revolutions, erosion can reach up to about 77~m in the most active, southern regions in our study. 
In contrast, it does not exceed $\sim$30~m for most northern features. 
We see that facets directed toward the equator, while in the northern hemisphere, sustain enhanced erosion compared to other facets at the same latitude. For those, it can reach the same level of erosion as is seen in the southern hemisphere. 
As a consequence of the trends in the surface energy distribution described in the previous section, the pattern of latitudinal variations for erosion is clearly observed. Indeed, the amount of erosion after ten orbits decreases when facets are located closer to the north pole. In the northern hemisphere, facets sustaining the most erosion are those closest to the equator, or perpendicular to the equatorial plane, as they receive direct insolation around successive perihelion passages. 
Nonetheless, some of these first order latitudinal effects are mitigated in part, due to the complex topographic shape of 67P's nucleus, which induces local self-heating.
The important result for these general considerations is that the amount of erosion achieved after ten orbits (the assumed current period of time that the comet has spent in its current orbit) never reaches the observed dimensions of any of the observed morphological structures. For example, the smallest feature in our study (feature 12, also known as Ma'at\_01) has a typical average size of $\sim$130~m, and would sustain an increase of its diameter by only 10 to 15~m  after ten orbits. The largest amount of erosion among our 380 facets remains below 80~m. Therefore, we confirm that pits cannot form by the progressive erosion of 67P's surface.

\subsubsection{Local variations}

\begin{figure*}[ht]
        \centering
        \includegraphics[width=\textwidth]{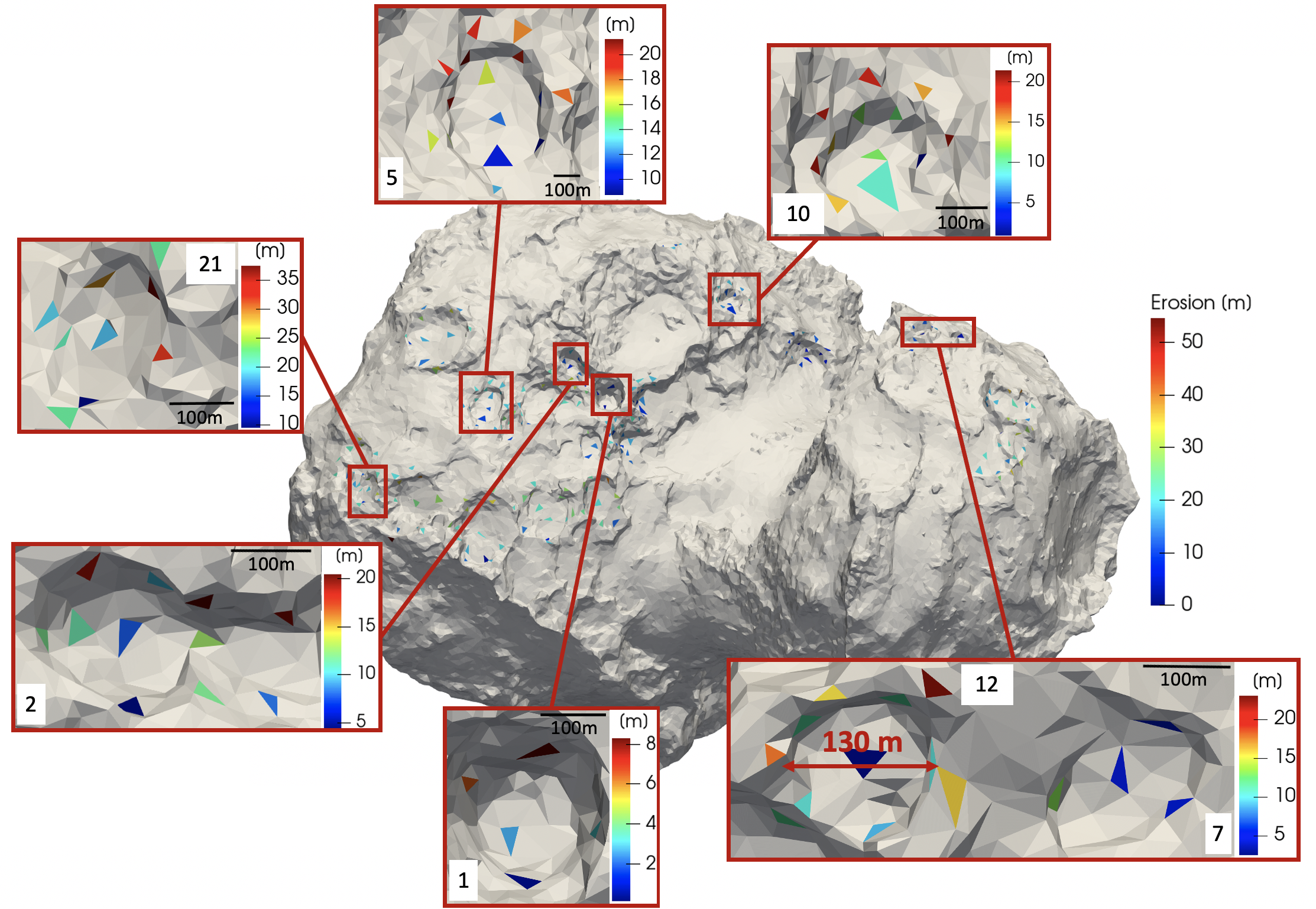} 
        \caption{Local examples of erosion achieved after ten orbits, highlighting differential erosion and flattening trends.}
        \label{local_er_3d}
    \end{figure*}

To the first order, the latitudinal pattern of erosion dominates. However, at the scale of each morphological structure, local trends appear similar across the surface. The first trend we observe is that erosion is generally more intense on the plateaus surrounding the pits when they are exposed to the Sun. 
In contrast, the bottoms of these pits do not sustain as much erosion, even after ten full orbital revolutions. This is especially true for circular pits with a high depth/diameter ratio (e.g., pits 1 or 2, also known as Seth\_01 and Seth\_02 with Seth\_03 combined, respectively, and pit 12, also known as Ma'at\_01). This general behavior tends to erase the local topography and leads to shallower features, such as those observed in the southern hemisphere.
In general, the walls of the pits experience some differential processing, with erosion enhanced along a specific direction (e.g., features 1 or 5 in Fig.~\ref{local_er_3d}). This is directly related to the asymmetric distribution of the input energy, especially when some facets receive direct insolation while others mostly get a self-heating contribution. This suggests that, if we account for the shape evolution due to erosion, elongated pits are more thermally processed than small circular ones. 
As a consequence, our results are consistent with deep, circular, or quasi-circular pits, such as the pits labeled as 1 (Seth\_01 on the big lobe) and 12 (Ma'at\_01 on the small lobe) in our study, which are the least processed pits, or the best preserved under the current illumination conditions. 

\section{Discussion} \label{sec:discussion}

\subsection{Local and global shape effects}

Our results show that the local topography and the complex global shape of the nucleus can considerably impact the energy balance at the surface (Sect. \ref{sec:shape_effects}). This is particularly true when considering the different sides of a given pit. As a result, some walls and bottoms of pits are not as easily reached by insolation as the corresponding exposed plateaus, making the onset of activity in the inner parts of these morphological features more difficult. 
It is thus necessary to take into account the effects of both shadowing and self-heating at the scale of these depressions. These processes are also important at the scale of 67P's nucleus, because its specific bilobate shape leads to the neck region being highly shadowed during the northern day. While self-heating is found to be mostly negligible compared to direct insolation for most facets we studied, it can be an important energy source in some cases, especially at the bottom of pits and around the neck region, where direct insolation is limited. In such locations, the contribution of self-heating to the local energy balance can reach up to 60\%.
These results are consistent with earlier studies. For instance, \citet{keller2015} showed that self-heating could reach 50\% of the total energy received in some areas of the neck region. \cite{macher2019} also showed that, even though the average contribution of self-heating in the regions they studied was evaluated to be 1\% of the direct insolation, it can be enhanced in rough areas not reached by direct insolation. In these locations, it could reach as much as 50\% of the direct insolation contribution. The important contribution of self-heating was also emphasized by \citet{tosi2019}, for deriving the temperature map at high spatial resolution ($<$15m) from the Visible InfraRed Thermal Imaging Spectrometer (VIRTIS-M) data.
The aforementioned studies were performed using various resolutions of 67P's shape model, which suggests that shadowing and self-heating are important at all scales. Therefore, our results are not very sensitive to the choice of spatial resolution for the shape model: using the 125k-facets shape model, with an average distance between facets' nodes of about 20~m \citep{marshall2018}, allows morphological features to be sampled without increasing the computation time required if smaller facets are chosen.
Overall, detailed knowledge of the energy balance at the surface on a local scale is thus a necessary condition to quantify the effect of thermally induced processes on the evolution of the cometary surface. However, as discussed below, we find that it is not sufficient to understand the evolution of the surface, since the energy input does not translate into phase transitions and erosion in a straightforward manner.

\subsection{Nonuniform properties}

We show that current illumination conditions cannot result in the formation of the deep circular pits with such characteristics as observed by \emph{Rosetta}. In the southern regions, where sublimation-driven erosion is the most effective, erosion reaches $\sim$80~m at best (Fig.~\ref{er_2d}). We now discuss how the choice of initial parameters used in our thermal evolution model may influence this outcome. For instance, an increased porosity could result in larger amounts of erosion, as much as 50\% for facets that receive the most energy (Sect. \ref{sec:porosity}). However, it is unlikely that the bulk material in the uppermost layers has a porosity greater than 75\% \citep{ciarletti2015}. An increased dust-to-ice mass ratio had a similar effect in our simulation outcomes (Sect. \ref{sec:dusttoice}), although we identified that this was actually more due to the resulting increase in thermal conductivity than the composition itself. Therefore, local variations in composition or thermophysical properties could also induce different amounts of local erosion. Such local heterogeneities have indeed been identified at the surface of 67P, with a spatial scale of tens of meters, sometimes associated with the local exposure of volatile ices \citep[e.g.,][]{filacchione2016, fornasier2016}. On a global scale, differences between the small and the big lobes have been inferred from variations in their mechanical properties \citep{el-maarry2016}, and physical characteristics. For instance, the small lobe has larger goose-bump features, fewer morphological changes, and less frequent and smaller frost areas than the big lobe \citep{fornasier2021}. From these, the authors inferred that the small lobe might have a lower volatile content than the big lobe. Instead, we chose to apply a uniform set of initial parameters. Thus, our erosion rates could vary if we accounted for the actual heterogeneity of the nucleus. Based on the outcomes of simulations performed to select this set of initial parameters in Sect. \ref{initial-params}, we can estimate that the final erosion would change by about 20\% at most, due to local changes of porosity, composition, or thermal properties, as observed by the suite of instruments on board \emph{Rosetta}. Nonetheless, our general trends, based on the relative erosion between plateaus and bottoms, and differential erosion, are not sensitive to these initial conditions. 
As a consequence, our quantitative study validates the qualitative trend suggested by \citet{vincent2017} that sublimation-driven erosion leads to shallower and larger depressions, effectively erasing sharp geological features with time.

\subsection{Dust mantle}

The presence of a dust layer thicker than $\sim$10~cm was able to quench the activity of most of the facets that we have studied (Sect. \ref{sec:mantle}). If we consider that a thick ($>$10~cm) dust layer was initially present throughout the surface of 67P when it arrived on its current orbit in 1959, we would find erosion rates lower than those obtained in our simulations. 
It is interesting to note from Fig.~\ref{mantle_3d_10cm_4outputs} that, when the appropriate conditions are met, the activity of some facets is such that an initially thick dust mantle can be removed after several perihelion passages. In our simulations, thinner dust mantles are indeed periodically removed and formed as a direct consequence of ice sublimation and gas drag of dust particles. Evidence for such a cyclic formation and removal of dust with the seasons, and for fallback material, has been reported \citep[e.g.,][]{thomas2015, attree2019}: a thickness of about 5 mm in northern regions has been reported \citep{herny2021b}. Through thermal evolution modeling, \citet{davidsson2022} obtained a resulting dust mantle typically thinner than 2~cm. 
On a local scale, dust mantles may play a significant role. They would additionally be affected by the heterogeneous gravitational potential, impacting the local dust deposition at the surface. High-resolution observations by \emph{Rosetta}/OSIRIS show that the bottom of deep, circular pits is relatively flat, and covered with a fine dust layer \citep[e.g., features 1 and 2, also known as Seth\_01, and Seth\_02 and Seth\_03, or feature 12, also known as Ma'at\_01;][]{sierks2015}. Some pits have boulders of various size on their floor, which \citet{vincent2015a} used as an indication of the erosion age of these structures. For instance, the authors suggested that the boulder-free floor of Ma'at\_01 could represent the least eroded pit, while Ma'at\_02 (feature 7 in our study) and Ma'at\_03 would be increasingly eroded, with degraded walls and accumulated material within the pits. Our simulations cannot account for such effects. However, the degradation of walls after they are weakened and the accumulation of wall material at the bottom essentially lead to the same trend we found: pits become larger and shallower with time.

\subsection{Active pits}

Our study thus supports the hypothesis, initially made by \citet{vincent2015a}, that the deep, circular pits are less processed (or better preserved) than the large or elongated ones. 
Interestingly, the more preserved features have been unambiguously revealed as the source of thin dust jets, arising from the edges of these depressions, which indicates that activity and erosion are currently occurring \citep{sierks2015}. More generally, \citet[][]{vincent2015a} identified two trends in the depth-to-diameter ratio (d/D) of pits at the surface of 67P: active pits have a high d/D (>0.3), while pits with no observed activity have a much smaller d/D.

From our results, we cannot exclude that large, relatively shallow pits could be active, as erosion is efficiently erasing the structures, especially in the southern regions (Sect. \ref{sec:thermal-simu}). Furthermore, these features typically receive high amounts of energy (integrated over an orbit, or at perihelion), such that adding moderately to highly volatile species triggered numerical instabilities in our simulations (Sect. \ref{sec:abundance}). The sublimation of CO and CO$_2$ for these facets might actually trigger some outbursts, which no model can simulate, as the process is highly nonlinear. Indeed, these numerically unstable facets are typically found in areas of the pits that sustained the most erosion in previous tests (i.e., parts that received the most energy), either close to perihelion or integrated across the orbit.
It is thus likely that these reflect bursts of activity driven by the sublimation of such species. However, the fact that no active outbursts were observed from these pits suggests that the sublimation fronts could actually be located deeper in the nucleus than in our model, after the insertion orbits. Moreover, the numerical instabilities were found within the first few orbital revolutions of 67P under current illumination conditions, and were thus not reflective of the time of the \emph{Rosetta} observations.

For the best preserved structures, facets never experience such dramatic numerical behavior, yet \emph{Rosetta} observations suggest some outbursts of activity. In our simulations, when volatile species are added, sublimation fronts slowly progress under the surface (Fig.~\ref{volatiles_fronts}), yet continue to contribute to the activity. It is therefore very likely that CO and CO$_2$ remain close to the surface in these geological features, in accordance with their relatively unaltered nature. A further interesting aspect is that, when adding volatile species, facets that remain active (vs. those whose activity is quenched by progressive dust mantling) are those located at the edges of pits (Fig.~8). This corresponds to the observed activity of these pits \citep{vincent2015a}. Therefore, our results  support the hypothesis that these morphological features are probably very well preserved, or are the least altered ones.
Even without these additional volatiles, however, water ice is able to sublimate preferentially from the walls rather than the bottoms.

\subsection{Implications for the evolution of pits}

We have shown that cometary activity tends to erase surface features, so that deep, circular pits are likely the least processed morphological structures on the surface (Sect. \ref{sec:erosion}).
Clearly, these pits could not have been formed by sublimation-driven erosion. We have investigated very different illumination conditions across 67P's surface. Under these conditions, the patterns of differential erosion, and the preference for eroding plateaus rather than bottoms of pits, are maintained. Therefore, we can extrapolate that different illumination conditions on a different orbit would have led to similar trends. Furthermore, even if the southern hemisphere is obviously more processed than the northern hemisphere, traces of larger depressions can be found, and there is no clear dependence of the distribution of depressions on latitude \citep{vincent2015a}. We can thus argue that pits were initially present on a global scale, and they likely evolved due to sublimation-driven erosion at various degrees on the surface of 67P.

Our results provide a quantitative confirmation for several studies since no quantification of the erosion through all the recent orbits has been performed before. Concerning the formation of pits, \citet{ip2016} performed a morphological and dynamical study, by which they found that pits on JFCs were likely formed prior to acquiring their current orbital elements. \citet{mousis2015} tested the formation of pits with three phase transitions (sublimation, amorphous water ice crystallization, and clathrate destabilization) and found that each of these processes would require a period of time much longer than the time spent by the comet in the inner solar system to form the observed pits.
\citet{guilbert-lepoutre2016} also attested that it is very unlikely for 200-m pits to form under current illumination conditions. 
Such conditions are, however, prone to the formation of smaller-scale geological features, such as shallow depressions of several meters in depth, probably formed due to progressive seasonal erosion \citep{bouquety2021,bouquety2021a}.
When it comes to the evolution of pits, \citet{belton2010} proposed an evolutionary sequence where pits are erased through cometary activity: initially found as acute depressions seen on 81P/Wild~2, they would progressively become shallower depressions as observed on 103P/Hartley~2, which is relatively older in terms of the sublimation process \citep{ip2016}.
\citet{vincent2017} studied the global topography of comets observed by spacecrafts and reaffirmed this trend.
This paper provides a quantification of the erosion rates sustained at the level of the pits during all the time that 67P spent as a JFC in the inner solar system, which vigorously reaffirms the previous studies, at least for 67P. \\ 

In the future, we will confront the trends established in our study by constraining the sublimation-driven erosion sustained by other cometary nuclei where pits have also been observed, in particular 103P/Hartley, \citep{syal2013}, 81P/Wild 2, \citep{brownlee2004}, and 9P/Tempel 1, \citep{thomas2013}. When it comes to understanding the origin of pits, we argue that feature 1 (Seth\_01), on the big lobe, and 12 (Ma'at\_01), on the small lobe, are the least processed. Notwithstanding local heterogeneity giving rise to various pit sizes, these features are thus likely representative of pits as they were formed. This needs to be kept in mind when we seek to constrain the thermal or physical processes that carve these structures, and which remain to be identified: any process invoked needs to be able to excavate a significant volume of material in a quasi-circular shape.


\section{Summary}

We have investigated the erosion of morphological features at the surface of 67P/Churyumov-Gerasimenko, dominated by water ice-driven sublimation. We selected 380 facets of a medium resolution shape model of the nucleus, sampling 30 pits and alcoves across the surface. The energy balance at the surface was then computed with a high temporal resolution, and by including shadowing and self-heating contributions. We then applied a thermal evolution model to quantify the amount of erosion sustained after ten orbital revolutions under current illumination conditions. 

Our study shows that a detailed knowledge of the energy balance at the surface on a local scale is a necessary condition to quantify the effect of thermally induced processes, but is not sufficient. Indeed, the energy input does not translate into phase transitions and erosion in a straightforward manner. Also, although seasons drive the global erosion trends, local topography can play a significant role in the final erosion state.

The erosional behavior on the surface revealed that morphological features such as pits and alcoves become larger and shallower with time: they are effectively erased through sustained cometary activity.
    
Finally, none of the surface structures we studied can be formed through progressive erosion. Pits Seth\_01 and Ma'at\_01 are among the least processed representatives of what pits would have looked like when they were formed, although the forming mechanism remains to be elucidated.


\begin{acknowledgements}
We thank the referee for their thorough review and comments which improved this paper. This study is part of a project that has received funding from the European Research Council (ERC) under the European Union’s Horizon 2020 research and innovation programme (Grant agreement No. 802699). We gratefully acknowledge support from the PSMN (Pôle Scientifique de Modélisation Numérique) of the ENS de Lyon for the computing resources. We thank the European Space Research and Technology Centre (ESTEC) and the European Space Astronomy Centre (ESAC) faculty council for supporting this research. JL thanks the support by the CNES. AB thanks the Swedish National Space Agency (SNSA) grant 108/18 for its support.
\end{acknowledgements}


\bibliographystyle{aa.bst}
\bibliography{Pits_Paper.bib}


\begin{appendix}

\section{Additional information}
    \begin{center}
    \begin{table}[!h]
    \centering
        \caption{Location and characteristics of the 30 pits.}   
        \begin{tabular}{c|rrccrcc}
         \hline
         ID & Lat & Lon & R$_n$ & D & d & Hemis- & Lobe\\  
            & [\textdegree] & [\textdegree] & [km] & [m] & [m] & phere  &     \\
         \hline \hline
        1  & 61 & -160 & 0.936 & 210 &150 &N &big\\
        2  & 53 & -159 & 1.262 & 150 &90 &N &big\\
        3  & 48 & -152 & 1.401 &175 &130 &N &big\\
        4  & 25  & -20  &2.262  &190 &55 &N &small\\ 
        5  & 35 & -153 & 1.693 &210 &60 &N&big\\
        6  & 37 & -149 & 1.574 &165 &85 &N&big\\  
        7  & 36 & 10 & 1.930 &155 &50 &N&small\\  
        8  & -12 & 109 & 1.570 &505 &85 &S&big\\
        9  & 24 & 63 & 1.128 &265 &95 &N&big\\
        10  & 53 & 89 & 1.203 &230 &70 &N&big\\
        11  & 26 & -14 & 2.301 &185 &60 &N&small\\ 
        12  & 45 & 5 & 1.810 &130 &60 &N&small\\ 
        13  & 15 & -135 & 1.758 &370 &120 &N&big\\
        14  & 19 & -129 & 1.414 &345 &80 &N&big\\
        15  & 25 & -123 & 1.085 &240 &85 &N&big\\  
        16  & 48 & -133 & 1.064 &135 &50 &N&big\\
        17  & 29 & -10 & 2.314 &205 &55 &N&small\\
        18  & 53 & -139 & 0.950 &265 &165 &N&big\\
        19  & 64 & -154 & 0.967 &220 &125 &N&big\\
        20  & 16 & -146 & 2.137 &275 &105 &N&big\\
        21  & 17 & -148 & 2.284 &140 &40 &N&big\\
        22  & 18 & -154 & 2.350 &210 &35 &N&big\\
        23  & 37 & -167 & 2.072 &380 &115 &N&big\\
        24  & -36 & 121 & 1.353 &355 &80 &S&big\\
        25  & -36 & 167 & 1.463 &685 &80 &S&big\\ 
        26  & 30 & 143 & 1.900 &655 &90 &N&big\\
        27  & 16 & 109 & 1.660 &290 &90 &N&big\\
        28  & -63 & -143 & 1.602 &215 &100 &S&big\\ 
        29  & 65 & 122 & 1.548 &165 &60 &N&big\\
        30  & 14 & 99 & 1.589 &240 &80 &N&big\\
         \hline
        \end{tabular}
        \centering \small R$_n$: average distance to the center of mass of the shape model. D and d: approximate diameter and depth of the depression, respectively.
        \label{table:position_coord}
        \label{table_pos}
    \end{table}
    \end{center}

\begin{figure*}[ht]
    \centering
    \includegraphics[width=\textwidth]{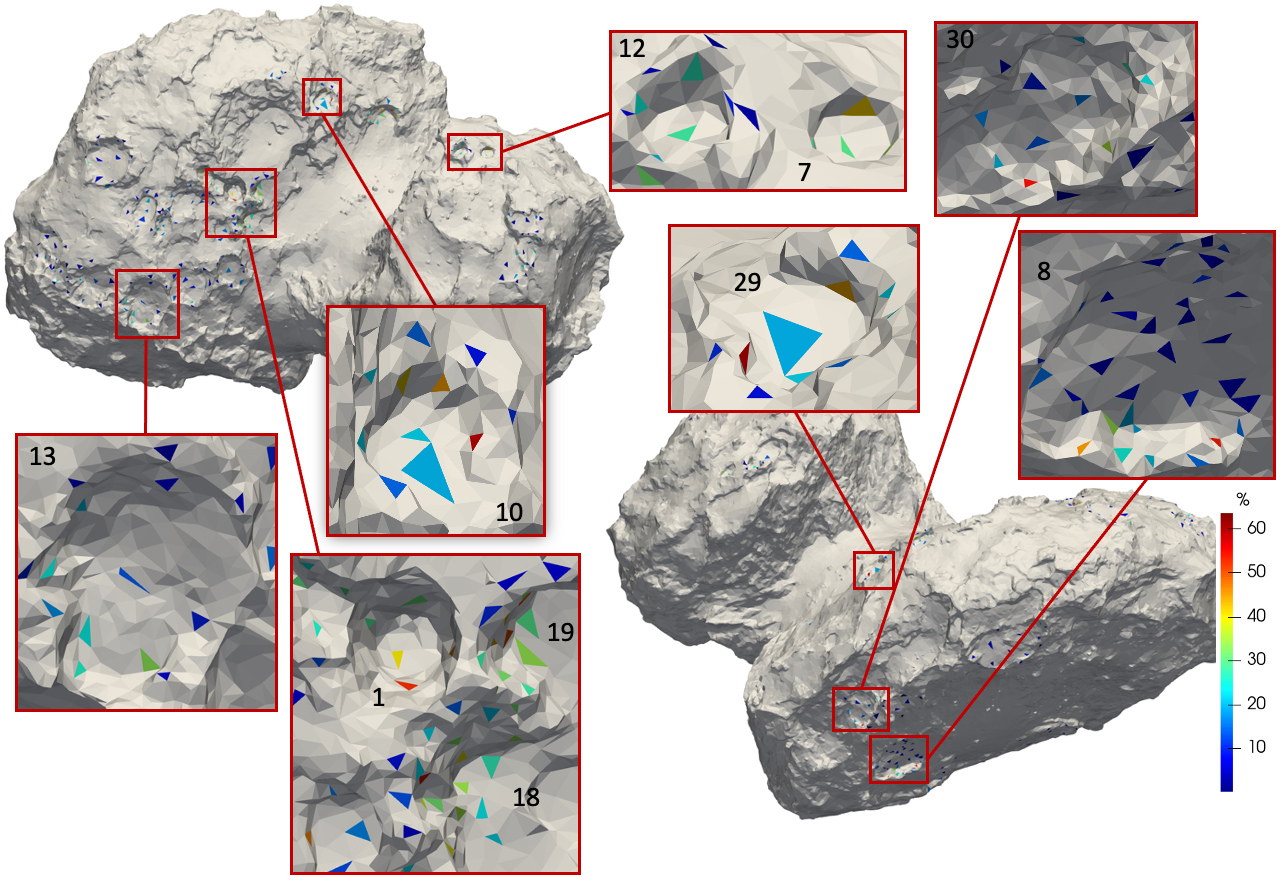}
    \caption{Fraction of the energy input from self-heating relative to the total energy received. We highlight some examples where the self-heating contribution is significant.} 
    \label{sh_3d}
\end{figure*}

\begin{figure*}[ht]
    \centering
    \includegraphics[width=\textwidth]{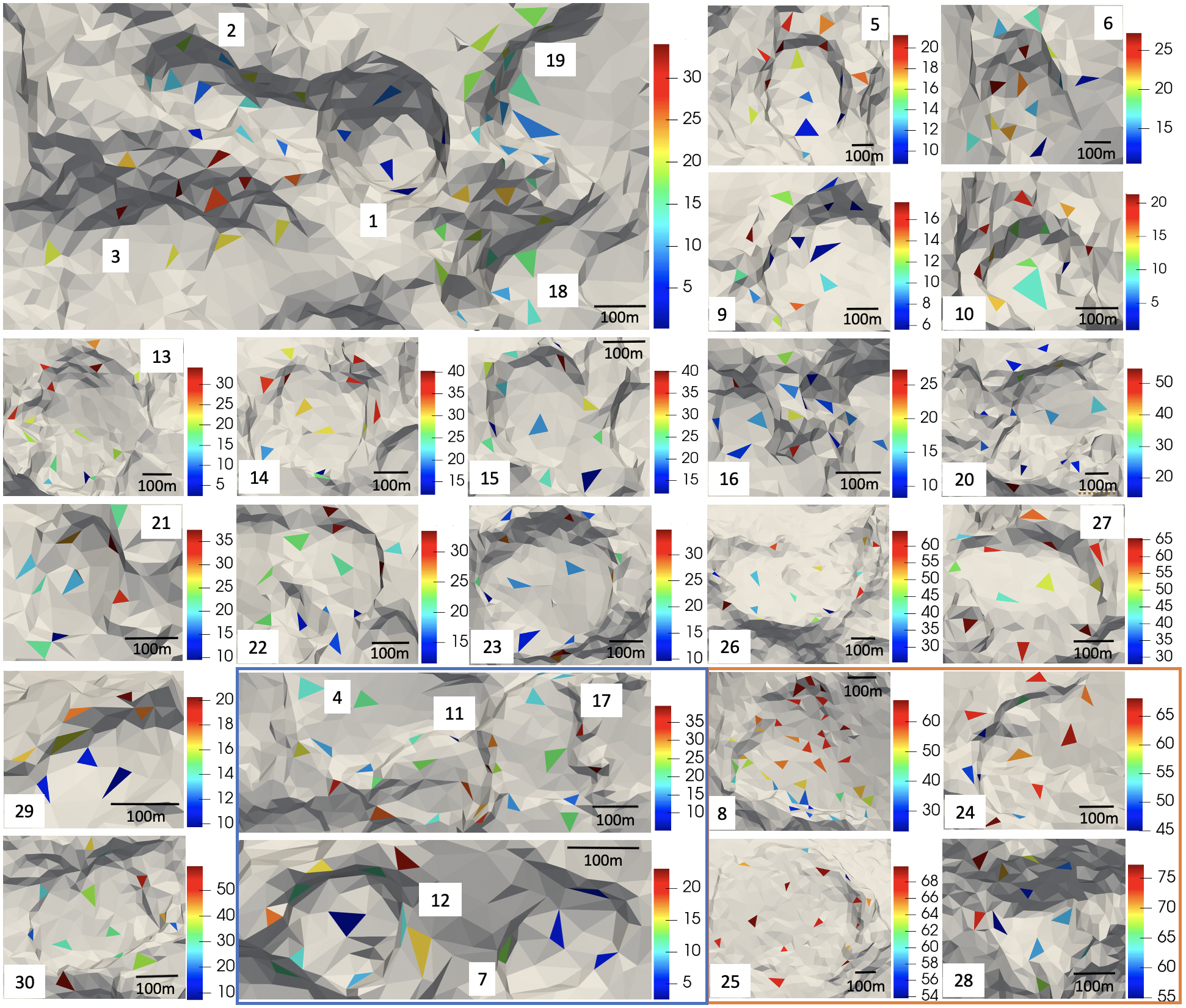}
    \caption{Erosion (in meters) achieved after ten orbital revolutions, for all facets and morphological features studied. The blue box contains depressions located in the small lobe, the orange box contains the big lobe's southern depressions, and the rest are located in the big lobe's northern hemisphere.} 
    \label{sh_3d}
\end{figure*}

\begin{figure*}[ht]
    \centering
    \includegraphics[width=12cm]{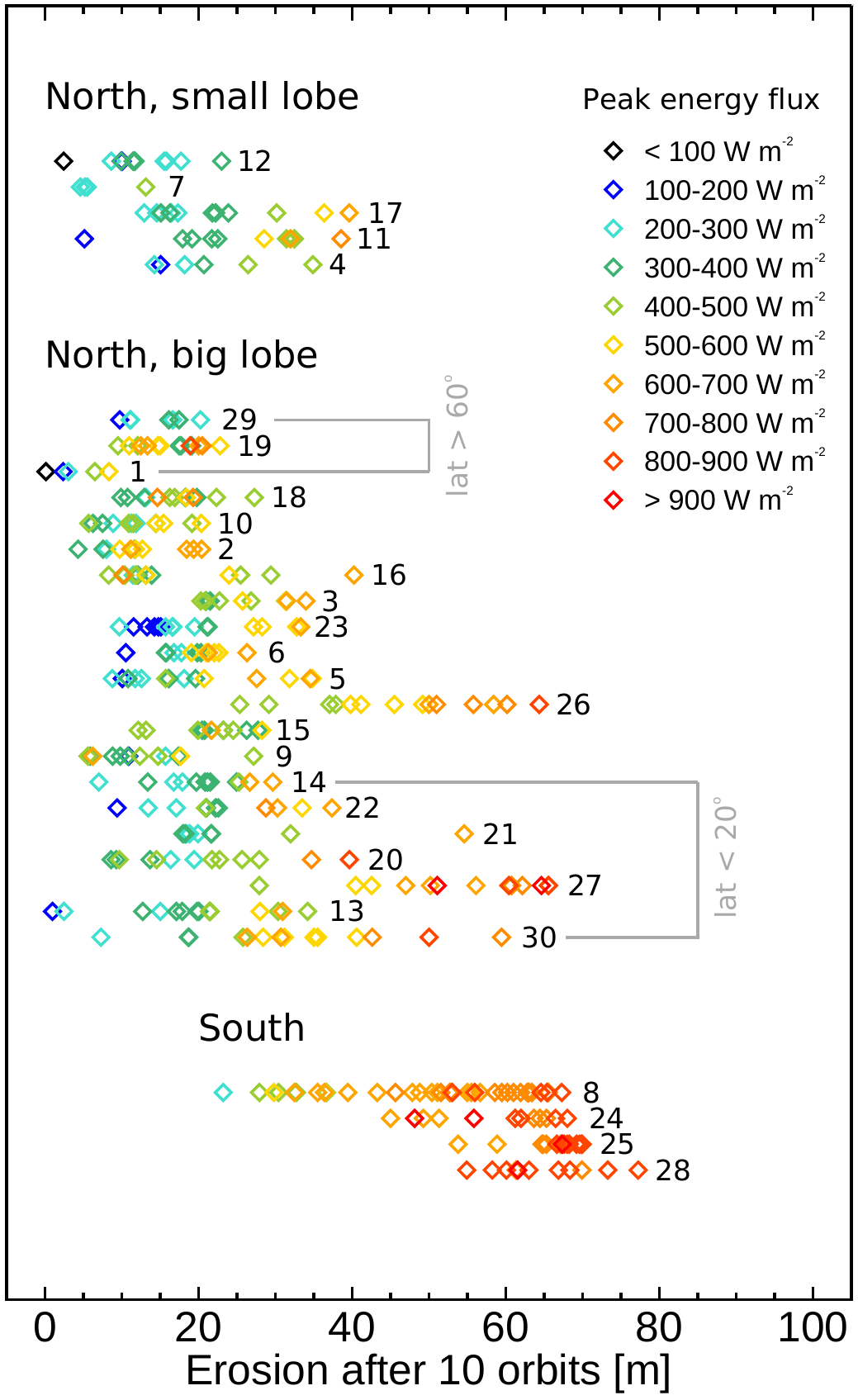}
    \caption{Erosion sustained after ten orbital revolutions for all the structures we studied. } 
    \label{laterosion}
\end{figure*}

\begin{figure*}[ht]
    \centering
    \includegraphics[width=18cm]{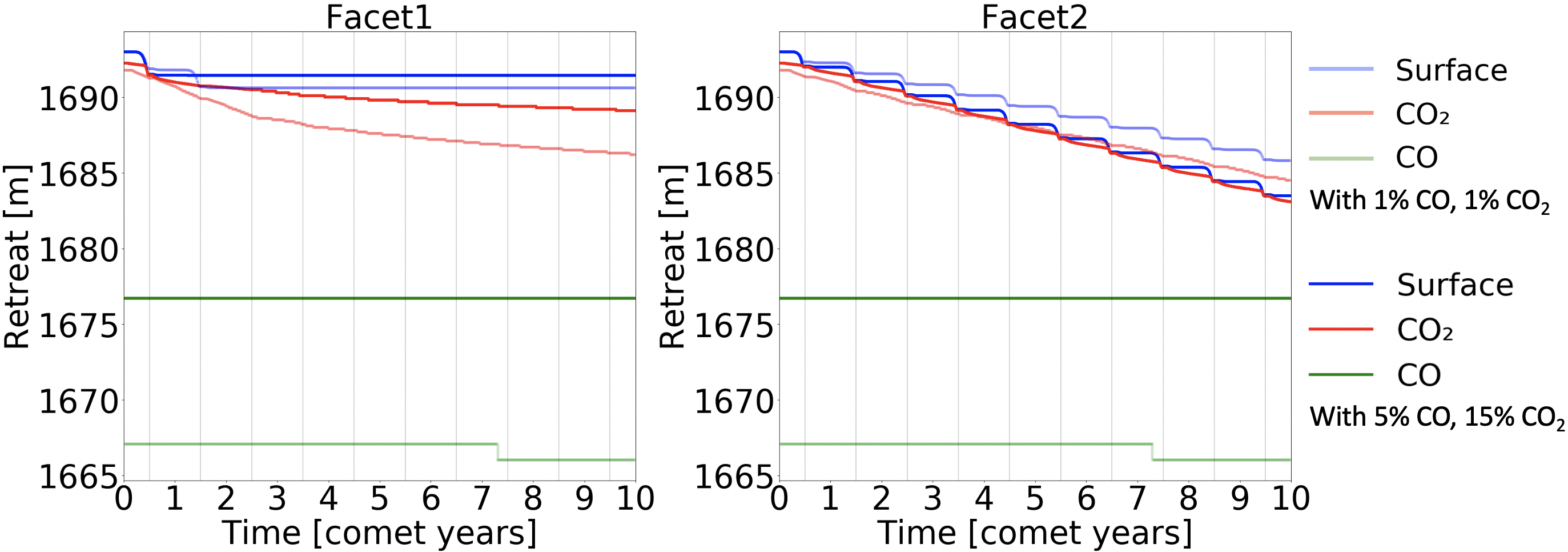}
    \caption{Subsurface retreat of the sublimation fronts of CO and CO$_2$ for the two facets studied in Sect. \ref{sec:abundance}. We display the two compositional cases: 1\% CO and CO$_2$, and 5\% CO and 15\% CO$_2$. The gray vertical lines mark the perihelions.}
    \label{volatiles_fronts}
\end{figure*}

\end{appendix}

\end{document}